\documentclass[prd,amsmath,amssymb,twocolumn]{revtex4-1}

\usepackage{graphicx}
\usepackage{color}
\usepackage{amsmath}
\usepackage{amssymb}

\usepackage{bm}
\usepackage{bbm}
\usepackage{cancel}
\usepackage{hyperref}

\newcommand{\CH}{\mathbb{C}\otimes\mathbb{H}}
\newcommand{\CO}{\mathbb{C}\otimes\mathbb{O}}
\newcommand{\CHO}{\mathbb{C}\otimes\mathbb{H}\otimes\mathbb{O}}
\newcommand{\RCHO}{\mathbb{R}\otimes\mathbb{C}\otimes\mathbb{H}\otimes\mathbb{O}}
\newcommand{\A}{\mathbb{A}}

\newcommand{\CLeight}{\mathbb{C}l(8)}

\newcommand{\CLsix}{\mathbb{C}l(6)}
\newcommand{\CLtwo}{\mathbb{C}l(2)}

\newcommand{\CLten}{\mathbb{C}l(10)}

\newcommand{\Gsm}{SU(3)$_\textup{C}\hspace{.3mm}\times\hspace{.3mm}$SU(2)$_\textup{L} \hspace{.3mm}\times\hspace{.3mm}$U(1)$_\textup{Y}$/$\hspace{.3mm}\mathbb{Z}_6$}
\newcommand{\gsm}{\mathfrak{su}(3)_C \oplus \mathfrak{su}(2)_L \oplus \mathfrak{u}(1)_Y}

\newcommand{\C}{\mathbb{C}}
\newcommand{\R}{\mathbb{R}}

\newcommand{\OHCR}{\mathbb{O}\oplus\mathbb{H}\oplus\mathbb{C}\oplus\mathbb{R}}

\usepackage{xcolor}

\begin{document}

\title{A Superalgebra Within: \\ representations of \\lightest standard model particles  \\form a $\mathbb{Z}_2^5$-graded algebra }

\author{ N. Furey}
\affiliation{$ $\\  Iris Adlershof, Humboldt-Universit\"{a}t zu Berlin,\\ Zum Grossen Windkanal 2,  Berlin, Germany  \\ and\\    AIMS South Africa, 6 Melrose Road, Muizenberg, Cape Town, South Africa\\and\\ Nicolaus Copernicus University, Grudziadzka 5, Toru\'{n}, Poland\vspace{1mm}\\ furey@physik.hu-berlin.de\\HU-EP-25/16}

\begin{abstract}
It is demonstrated how a set of particle representations, familiar from the Standard Model, collectively form a superalgebra.  Those representations mirroring the internal behaviour of the Standard Model's gauge bosons, and three generations of fermions, are each included in this algebra, with exception only to those irreps involving the top quark.  This superalgebra is isomorphic to the Euclidean Jordan algebra of $16\times 16$ hermitian matrices, $\mathcal{H}_{16}(\C),$ and is generated by division algebras.  The division algebraic substructure enables a natural factorization between internal and spacetime symmetries.  It also allows for the definition of a  $\mathbb{Z}_2^5$ grading on the algebra.  Those internal symmetries respecting this substructure are found to be $\gsm,$ in addition to four iterations of $\mathfrak{u}(1)$.  For spatial symmetries, one finds multiple copies of  $\mathfrak{so}(3)$.  Given its Jordan algebraic foundation, and its \it apparent \rm non-relativistic character, the model may supply a bridge between particle physics and quantum computing.


We close  by describing current research directions.  These include (1) detailing how this construction fits into the larger picture of \it Bott Periodic Particle Physics, \rm first introduced in \citep{Gen}, \citep{Fur2021}, \citep{Fur2023}, (2) investigating how the origin of this Peirce decomposition might be grounded in the unsung algebra $\OHCR$, and (3) proposing how these two directions may merge by reframing the model in terms of the 16$\hspace{.5mm}\mathbb{R}$ dimensional sedenion algebra, $\mathbb{S}$. \end{abstract}

\maketitle

This paper is written with the intention of being accessible to a broad range of physicists and mathematicians.  Its central mathematical structure is that of hermitian
matrices.

\section{A $\mathbb{Z}_2$ grading within}

High-energy physicists have long hoped that particle detectors would expose a certain 1-to-1 correspondence between the bosons and fermions of elementary particle physics, \citep{atlas}.  However, as of the time of this writing, the LHC has failed to confirm the prediction in its usual incarnation.  In this article, we propose that it may not be necessary to forfeit the idea in its entirety.  That is, instead of searching outside  the Standard Model for the desired $\mathbb{Z}_2$ grading, it may be possible to search within.

As with SUSY models, the model proposed in this article is based on the structure of a superalgebra.  However, we use this superalgebra in a different way, and as a result, find a description of elementary particles that is not supersymmetric. Notably, this description does not entail a doubling of the Standard Model's particles. 

It is rarely emphasized in the literature, if at all, that the ratio of the number of fermionic degrees of freedom to bosonic degrees of freedom in the Standard Model is approximately 3:1.  We observe that if one were to recast the third generation of fermions as a \it na\"{i}vely bosonic \rm product $\sim \Psi \Psi^{\dagger}$, then this ratio, incidentally, becomes 1:1.  

In this article, we do indeed identify a $\mathbb{Z}_2$ grading within the Standard Model's existing irreps, not only once, however,  but several times over.  This $\mathbb{Z}_2^5$ grading makes obvious a $\left(64+3\cdot 64\right)\hspace{.5mm}\mathbb{R}$ partitioning of the $256 \hspace{.5mm}\mathbb{R}$ dimensional $\mathcal{H}_{16}(\C).$  Each of the Standard Model's known irreps are included in this minimalistic superalgebra, with exception only to those representations involving the top quark.  As a result, one might speculate that the top quark is fully composite in nature, while the bottom quark is partially composite,~\citep{kaplan},  \citep{tqc}, \citep{topcomp}.  For a preview see Figure~\ref{map}.

As an alternative perspective, we argue that there may be a way to introduce all third generation states via a product, reminiscent of the spinor-helicity formalism.  This $\mathcal{H}_{16}(\C)$ superalgebra, after all, may be viewed as a direct extension of  $\mathcal{H}_{2}(\C),$ wherein familiar Weyl operators, $p_{\mu}\sigma^{\mu},$  reside.

As will be seen towards the end of this article, the spatial symmetries respecting the division algebraic $\mathbb{Z}_2^5$ grading lend themselves to an interpretation that matter particles are extended objects.  Intuitively, they are the ``square root" of $p_{\mu}\sigma^{\mu},$ or rather, the full momentum-space covariant derivative.  The model does not appear to require that these objects be embedded into a background spacetime.

\section{In Context}

Upon a first glance at the Standard Model's list of fundamental particles, it might be difficult to see much more than randomness.  Why such a long, and erratic sequence of irreducible representations?

However, patterns do unquestionably exist.  To be explicit, the Standard Model's fermions are known to come in three copies, or generations.  As mentioned above, its fermion-to-boson ratio is roughly 3:1.  Furthermore, we know that SU(2)$_L$ violates parity maximally.  Our failure to understand the bigger picture is not a proof that no such bigger picture exists.  

Upon the inclusion of three generations of right-handed neutrinos, the total number of local off-shell (unconstrained) degrees of freedom in the Standard Model comes to approximately $250 \hspace{.5mm}\mathbb{R},$ that is, depending on how the Higgs is included, and on how Weyl operators are handled.  Given a dearth of new particle discoveries at the LHC, it may be advisable to search for mathematical structures of this same size.  Intriguingly, a few candidates do readily emerge.  For example, the exceptional Lie algebra $\mathfrak{e}_8$ occupies $248 \hspace{.5mm}\mathbb{R},$ while the Clifford algebra $Cl(0,8)$ of Bott Periodicity occupies $256 \hspace{.5mm}\mathbb{R}.$ $Cl(0,8)$ may be viewed as a certain real slice of the complex Clifford algebra $\CLeight$.  The same can be said for the $256 \hspace{.5mm}\mathbb{R}$-dimensional $\mathcal{H}_{16}(\C)$, which we explore here.

Beyond having an appropriate capacity, one may also require some inner substructure that will prompt the symmetries of these algebras to break.  That is, if one opts not to rely upon arbitrarily chosen Higgs fields.  It is known that both $Cl(0,8)$ and $\mathfrak{e}_8$ are closely tied to composition algebras that, in one way or another, generate them.  The same can be said for $\mathcal{H}_{16}(\C)$.  

In the existing literature, there is no shortage of  models built to address the three-generation problem.  Notable proposals include, but of course are not limited to~\citep{Silagadze}-\citep{Gunnew}.  We point out in particular a long-running interest in the 27-dimensional exceptional Jordan algebra, $\mathcal{H}_3(\mathbb{O}),$~\citep{Silagadze}, \citep{oct_e6}, \citep{DV}, \citep{DVtod},   \citep{boyle1}, \citep{Jac2021}, \citep{Per2021}, \citep{singh_v}.  (In future work, we will examine $\mathcal{H}_3(\CHO),$ the 3$\times$3 hermitian matrices over $\CHO$.)  Also closely related are $\mathfrak{e}_8$ models, advocated for in~\citep{Ram1977}, \citep{Che2020}, \citep{Man2022}, \citep{TTT}.

In this article, we will generate $\mathcal{H}_{16}(\C)$ from the left action of the algebra $\CHO$ on itself.  There is a long history of authors making use of $\CHO$ in various ways, starting with Dixon.  See for example, \citep{Dix2004}, \citep{Dixon_recent}, \citep{Gen}, \citep{thesis}, \citep{321}, \citep{Carlos2019}, \citep{mg16}, \citep{fh1}, \citep{dasb}, \citep{Fur2021}, \citep{mosaic}, \citep{FR2022}, \citep{Che2023}, \citep{Fur2023}, \citep{Jens2023}, \citep{ncgcliff}, \citep{TTT}.

Of special relevance is a long-running research programme of Alvarez, Delage, Valenzuela, and Zanelli, known as \emph{Unconventional SUSY}, \citep{z_ped}, \citep{z_chiral}.  There, a related superconnection is proposed,  in the form of a Lie superalgebra. 

Also in close connection is the work of Balbino, de Freitas,  Rana,  Toppan, \citep{toppan}, Stoilova and Van der Jeugt, \citep{jug}, Aizawa, Kuznetsova, Toppan \citep{aiz}, and colleagues in the realm of $\mathbb{Z}_2^n$ graded algebras and their applications in physics.

This article is a distillation of a long-running research effort.  Early three-generation results appeared in~\citep{Gen}, \citep{thesis}, \citep{321}, based on the Clifford algebra $\CLsix$, generated by the left action of the complex octonions on themselves.  Already in \citep{Gen} a proposal was put forward to extend to $\CLeight$ by including the quaternions.  From here, it was suggested to take some real slice of $\CLeight$, be it as the Bott Periodic $Cl(0,8),$ e.g. \citep{Fur2021}, \citep{dasb}, \citep{Fur2023}, or as the hermitian slice $\mathcal{H}_{16}(\C),$ e.g. \citep{mg16}, \citep{dasb},\citep{mosaic}, \citep{FR2022}.  Numerous seminars may be found online.

\section{Content\label{toc}}

In Section~\ref{particlecontent} we describe the Standard Model's particle content.  In Section~\ref{JAs} we introduce the physically relevant Euclidean Jordan algebras $\mathcal{H}_2(\C)$ and $\mathcal{H}_{16}(\C)$.     In Section~\ref{HLA} we describe how $\mathcal{H}_{16}(\C)$ can be generated by $\CHO$, via its left multiplication algebra, $L_{\CHO}$, which we define.  

In Section~\ref{early} we summarize early three-generation results.  

In Section~\ref{assemble}, we define a Peirce decomposition of $\mathcal{H}_{16}(\C)$, based on division algebraic substructure.   Symmetries respecting this division algebraic substructure are then identified.  Surviving octonionic symmetries are found to be $\mathfrak{u}(3)\oplus\mathfrak{u}(2)\oplus 3\mathfrak{u}(1),$ wherein lies the Standard Model's internal gauge symmetries $\gsm.$  Non-redundant quaternionic symmetries are found to be $\mathfrak{so}(3)$, and are expected to eventually give rise to spatial symmetry, although in an unexpected way.  Anomaly cancellation will need to be addressed in future work.

Applying these symmetries to $\mathcal{H}_{16}(\C)$ leads to an interpretation of the algebra as a space of \it Weyl super operators. \rm We find a (cursory) momentum-space covariant derivative of dimension $64\hspace{.5mm}\mathbb{R},$ and three generations, each of dimension $64\hspace{.5mm}\mathbb{R}.$  There is no overlap between these spaces, so that one finds $\mathcal{H}_{16}(\C)$ partitioned as $64\hspace{.5mm}\mathbb{R} + 3\cdot 64\hspace{.5mm}\mathbb{R} = 256\hspace{.5mm}\mathbb{R}.$  The third generation lacks only those representations involving the top quark.  

In this section, we furthermore propose an alternative product description for the third generation, motivated by Penrose's early work, and a recent article on triality~\citep{TTT}.

We identify the Weyl operators $p_{\mu}\sigma^{\mu}$ as possible \it dark matter candidates. \rm

In Section~\ref{Z2} we demonstrate how $\mathcal{H}_{16}(\C)$ and our Peirce decomposition induce a $\mathbb{Z}_2$ grading, hence forming a superalgebra from the Standard Model's local degrees of freedom.  We then show how the grading can in fact be extended to $\mathbb{Z}_2^5.$  As one special case, we demonstrate how multiplication yields the action of the covariant derivative on fermion states.  As another special case, representations corresponding to gluons and W bosons are formed by multiplying quark representations together. 

In Section~\ref{PVM}, we identify sets of Peirce idempotents as  projective measurements.  Furthermore, we identify a certain \it algebra of observables \rm as the algebra of real linear combinations of Peirce idempotents.  Of special interest for us is the fact that the Peirce idempotents defined in this article \it preclude the observation of colour. \rm

In Section~\ref{disc}, we point out that spatial $\mathfrak{so}(3) = \mathfrak{su}(2)$ symmetries arise in precisely the same way that internal symmetries do.  Perhaps this could have meaningful implications for quantum gravity, e.g. \citep{Ashtekar}.   We mention that even though Lorentzian symmetry does not arise as a unitary symmetry on the Jordan algebra, it does appear nonetheless since the proper orthochronous group SO$^+$(3,1) constitutes the non-trivial inner automorphisms of $\CH.$  It is this group of inner automorphisms, in fact, that establishes a complementary relationship between generations (mass) and spin in the model.

The existence of an independent $\mathfrak{so}(3)$ symmetry for each diagonal Peirce block leads to an interpretation that fermions be understood as extended objects.    However, with this said, one finds no need to introduce a background spacetime for this model.  

It is pointed out that the set of all fermion representations in this model may be represented by the complete graph $K_5$.  It is proposed that this feature, in addition to Jordan algebraic structure, may eventually enable this model of particle physics to be understood in terms of quantum computing.  Please see~\citep{Hilary}, \citep{BoyleKulp}, \citep{LiBoyle}, \citep{Fotini}, \citep{seth}, \citep{MarCos}.

After the conclusion, we describe areas of current development.  In Section~\ref{Bigpic} we describe how extending to $\CHO$'s full multiplication algebra allows this current work to fit into a wider research program known as Bott Periodic Particle Physics, \citep{Fur2021}, \citep{Fur2023}.  It also opens the door to a description of gravity via generalized tetrads.   In Section~\ref{plus}, we propose that the Peirce decomposition seen in Figure~\ref{map} may ultimately stem from the little-known 15$\hspace{.5mm}\R$ dimensional algebra $\OHCR,$ or further,  the nested Cayley-Dickson embeddings $\mathbb{R}\subset \mathbb{C}\subset \mathbb{H}\subset\mathbb{O}\subset\mathbb{S}.$

\section{The Standard Model's Particle content \label{particlecontent}}   

The Standard Model is based on  the gauge group \Gsm, and  spacetime symmetries of SL(2,$\C$), or more precisely, the Poincar\'{e} group with its Lorentz subgroup extended to the double cover.  Its degrees of freedom are said to reside within the vector spaces afforded by the  irreducible representations below.  

The  irreducible representations below are identified with their particle names in grey.  Their behaviour under  the SU(3) colour group is given by the first entry, their behaviour under the SU(2) weak isospin group is given in the second entry, weak hypercharge is given in the third entry, and their behaviour under SL(2,$\C)$ appears in the subscript.  (In keeping with the representation labeling of other groups, SL(2,$\C)$ Weyl spinors are described here as 2-dimensional.)  The descriptions below correspond to off-shell (unconstrained) representations.

\begin{center} ( \hspace{1mm}SU(3)$_\textup{C}$, \hspace{1mm} SU(2)$_\textup{L}$, \hspace{1mm} U(1)$_\textup{Y}$ \hspace{1mm})$_{\textup{SL}(2,\C)},$
\end{center}

\noindent Left-handed fermions (complex representations)
$$\begin{array}{llll} 
\color{gray}{(\hspace{1mm}  u, \hspace{1mm}d\hspace{1mm})_L} & \left( \hspace{1mm} \underline{\mathbf{3}},\hspace{1mm} \underline{\mathbf{2}}, \hspace{1mm}\frac{1}{6}\hspace{1mm} \right)_2 \hspace{6mm}   &  \color{gray} (\hspace{1mm}  \nu_e, \hspace{1mm}e\hspace{1mm})_L & \left( \hspace{1mm} \underline{\mathbf{1}},\hspace{1mm} \underline{\mathbf{2}}, \hspace{1mm}-\frac{1}{2}\hspace{1mm} \right)_2 \vspace{1mm}\\
\color{gray}(\hspace{1mm}  c, \hspace{1mm}s\hspace{1mm})_L & \left( \hspace{1mm} \underline{\mathbf{3}},\hspace{1mm} \underline{\mathbf{2}}, \hspace{1mm}\frac{1}{6}\hspace{1mm} \right)_2 \hspace{4mm}   &   \color{gray} (\hspace{1mm}  \nu_{\mu}, \hspace{1mm}\mu\hspace{1mm})_L & \left( \hspace{1mm} \underline{\mathbf{1}},\hspace{1mm} \underline{\mathbf{2}}, \hspace{1mm}-\frac{1}{2}\hspace{1mm} \right)_2  \vspace{1mm}\\ 
\color{gray}(\hspace{1mm}  t, \hspace{1mm}b\hspace{1mm})_L &\left( \hspace{1mm} \underline{\mathbf{3}},\hspace{1mm} \underline{\mathbf{2}}, \hspace{1mm}\frac{1}{6}\hspace{1mm} \right)_2 \hspace{4mm}   &\color{gray} (\hspace{1mm}  \nu_{\tau}, \hspace{1mm}\tau\hspace{1mm})_L & \left( \hspace{1mm} \underline{\mathbf{1}},\hspace{1mm} \underline{\mathbf{2}}, \hspace{1mm}-\frac{1}{2}\hspace{1mm} \right)_2  \vspace{1mm}\\
\end{array}$$

\noindent Right-handed fermions (complex representations)
$$\begin{array}{llllll} 
\color{gray}u_R & \left( \hspace{1mm} \underline{\mathbf{3}},\hspace{1mm} \underline{\mathbf{1}}, \hspace{1mm}\frac{2}{3}\hspace{1mm} \right)_2 \hspace{2mm} &  \color{gray}d_R & \left( \hspace{1mm} \underline{\mathbf{3}},\hspace{1mm} \underline{\mathbf{1}}, \hspace{1mm}-\frac{1}{3}\hspace{1mm} \right)_2 \hspace{2mm}  &  \color{gray}e_R &\left( \hspace{1mm} \underline{\mathbf{1}},\hspace{1mm} \underline{\mathbf{1}}, \hspace{1mm}-1\hspace{1mm} \right)_2 \hspace{2mm} \vspace{1mm} \\
\color{gray}c_R &\left( \hspace{1mm} \underline{\mathbf{3}},\hspace{1mm} \underline{\mathbf{1}}, \hspace{1mm}\frac{2}{3}\hspace{1mm} \right)_2    &  \color{gray}s_R &\left( \hspace{1mm} \underline{\mathbf{3}},\hspace{1mm} \underline{\mathbf{1}}, \hspace{1mm}-\frac{1}{3}\hspace{1mm} \right)_2   &  \color{gray}\mu_R &\left( \hspace{1mm} \underline{\mathbf{1}},\hspace{1mm} \underline{\mathbf{1}}, \hspace{1mm}-1\hspace{1mm} \right)_2  \vspace{1mm}  \\
\color{gray}t_R &\left( \hspace{1mm} \underline{\mathbf{3}},\hspace{1mm} \underline{\mathbf{1}}, \hspace{1mm}\frac{2}{3}\hspace{1mm} \right)_2   & \color{gray}b_R &\left( \hspace{1mm} \underline{\mathbf{3}},\hspace{1mm} \underline{\mathbf{1}}, \hspace{1mm}-\frac{1}{3}\hspace{1mm} \right)_2    &  \color{gray}\tau_R &\left( \hspace{1mm} \underline{\mathbf{1}},\hspace{1mm} \underline{\mathbf{1}}, \hspace{1mm}-1\hspace{1mm} \right)_2  \vspace{1mm}
\end{array}$$

\noindent Gauge bosons (real representations)
$$\begin{array}{llllll} 
\color{gray}G_{\mu} &\left( \hspace{1mm} \underline{\mathbf{8}},\hspace{1mm} \underline{\mathbf{1}}, \hspace{1mm}0\hspace{1mm} \right)_4  \vspace{1mm}  &\hspace{4mm}  \color{gray}W_{\mu} &\left( \hspace{1mm} \underline{\mathbf{1}},\hspace{1mm} \underline{\mathbf{3}}, \hspace{1mm}0\hspace{1mm} \right)_4  \vspace{1mm}  &  \hspace{4mm}\color{gray}B_{\mu} &\left( \hspace{1mm} \underline{\mathbf{1}},\hspace{1mm} \underline{\mathbf{1}}, \hspace{1mm}0\hspace{1mm} \right)_4 
\end{array}$$

\noindent Higgs (complex representation)
$$\begin{array}{ll} 
\color{gray}H &\left( \hspace{1mm} \underline{\mathbf{1}},\hspace{1mm} \underline{\mathbf{2}}, \hspace{1mm}-\frac{1}{2}\hspace{1mm} \right)_1.  
\end{array}$$

It is something of an enigma that in the decades since its inception, experiment has hardly nudged the Standard Model from its original particle content.  Notably in 2015 the Nobel prize was awarded to Super-Kamiokande and the Sudbury Neutrino Observatory for the discovery of neutrino oscillations.  The finding fares well with the possibility of the existence of right-handed neutrinos:  what would be a first, and frankly, instinctive correction to the theory's particle content.  The corresponding irreducible representations for these particles are found below.

\medskip

\noindent Sterile neutrinos (complex representations)
$$\begin{array}{lll} 
\color{gray}(\nu_e)_R \color{black}\hspace{1mm}\left( \hspace{1mm} \underline{\mathbf{1}},\hspace{1mm} \underline{\mathbf{1}}, \hspace{1mm}0\hspace{1mm} \right)_2&\hspace{1mm} \color{gray}(\nu_\mu)_R \color{black}\hspace{1mm}\left( \hspace{1mm} \underline{\mathbf{1}},\hspace{1mm} \underline{\mathbf{1}}, \hspace{1mm}0\hspace{1mm} \right)_2&\hspace{1mm}\color{gray}(\nu_\tau)_R\color{black} \hspace{1mm}\left( \hspace{1mm} \underline{\mathbf{1}},\hspace{1mm} \underline{\mathbf{1}}, \hspace{1mm}0\hspace{1mm} \right)_2
\end{array}$$

\section{Jordan algebras \\in terms of  \\outer products of spinors \label{JAs}}

A Jordan algebra $J$ is defined as a  non-associative algebra over a field $\mathbb{F}$, with multiplication  $\circ: J\times J\rightarrow J$ such that 
\begin{equation}
\begin{array}{lll}
(1)&\hspace{1mm}&J_1\circ J_2 = J_2\circ J_1,  \hspace{2mm} \textup{and}\vspace{2mm}\\
(2)&&(J_1\circ J_2)\circ (J_1\circ J_1) = J_1\circ (J_2\circ (J_1\circ J_1)) 
\end{array}
\end{equation}
\noindent $\forall J_1,J_2 \in J.$  From this definition, it would hardly be appreciated that Jordan algebras are intimately related to (also non-associative) Lie algebras in many physically relevant cases.

\subsection{$\mathcal{H}_2(\C)$\label{h2c}}

It is widely known that any  $2\times 2$ hermitian matrix, $p_{\alpha\dot{\beta}},$ may be written as a sum of outer products of 2-component complex Weyl spinors:
\begin{equation} p_{\alpha\dot{\beta}} = p_{\mu}(\sigma^{\mu})_{\alpha \dot{\beta}}=  \sum_{i=1}^2 \left(\psi_i \psi_i^{\dagger}\right)_{\alpha \dot{\beta}},
\end{equation} 
\noindent for $p_{\mu}\in\R,$ $\psi_i \in \C^2$, and $\sigma^{\mu}$ defined in the usual way:
\begin{equation}\begin{array}{ll}
\sigma^0:= \left( \begin{matrix}1&0 \vspace{1mm}\\ 0&1 \end{matrix}  \right), &
\hspace{2mm}\sigma^1:= \left( \begin{matrix}0&1 \vspace{1mm}\\1&0 \end{matrix}  \right),
\vspace{2mm} \\
\sigma^2:= \left( \begin{matrix}0&-i \vspace{1mm}\\ i&0 \end{matrix}  \right), &
\hspace{2mm}\sigma^3:= \left( \begin{matrix}1&0 \vspace{1mm}\\0&-1 \end{matrix}  \right).
\end{array}\end{equation}
\noindent It so happens that these $\psi_i,$ and consequently $p_{\mu} \sigma^{\mu},$ may be rewritten in terms of the nilpotent objects
\begin{equation}\label{alpha_matrix}
\beta:=  \left( \begin{matrix}0&1 \vspace{1mm}\\ 0&0 \end{matrix}  \right),\hspace{5mm} \beta^{\dagger}:=  \left( \begin{matrix}0&0 \vspace{1mm}\\ 1&0 \end{matrix}  \right).
\end{equation}
\noindent Explicitly, we may define an \emph{algebraic vacuum}, $v_0,$ as
\begin{equation} v_0:= \beta \beta^{\dagger} =  \left( \begin{matrix}1&0 \vspace{1mm}\\ 0&0 \end{matrix}  \right),
\end{equation}
\noindent and subsequently $\psi$ as
\begin{equation}   \psi := \psi^{\uparrow} v_0 + \psi^{\downarrow}\hspace{.5mm} \beta^{\dagger} v_0 =  \left( \begin{matrix}\psi^{\uparrow}&0 \vspace{1mm}\\ \psi^{\downarrow}&0 \end{matrix}  \right)
\end{equation} 
\noindent for $\psi^{\uparrow}, \psi^{\downarrow}\in\C.$  Consequently, we find hermitian matrices written purely in terms of ladder operators $\beta$ and $\beta^{\dagger}$ as
\begin{equation}\label{2ladder}
\sum_{\mu=1}^4p_{\mu} \sigma^{\mu} = c_{1} \hspace{.5mm}v_0   + c_{2} \hspace{.5mm}v_0 \beta +c_{3} \hspace{.5mm}\beta^{\dagger} v_0+ c_{4} \hspace{.5mm}\beta^{\dagger} v_0 \beta \hspace{1mm}+\hspace{1mm} h.c.,\vspace{2mm}\\
\end{equation} 
\noindent for $c_{i}\in\C,$ and where $h.c.$ is the hermitian conjugate.  Equation~(\ref{2ladder}) may finally be written more compactly as
\begin{equation}\label{outer}
\sum_{\mu=1}^4p_{\mu} \sigma^{\mu} = \sum_{i,j\in\{0,1\}} c_{ij} \hspace{.5mm}\beta^{\dagger}_iv_0 \beta_j \hspace{1mm}+\hspace{1mm} h.c.,\vspace{2mm}\\
\end{equation} 
\noindent where we have now defined $\beta_1:=\beta,$ and $\beta_0:=\mathbb{I},$ the identity.

It is well-known that the algebra of 2$\times$2 complex matrices, $M_2(\C),$ coincides with the complex Clifford algebra $\CLtwo,$ when thought of as a matrix algebra under multiplication $m(a,b)=ab$ for $a,b\in\CLtwo$.   With regard to the hermitian conjugate, $M_2(\C)\simeq \CLtwo$ splits into hermitian (Jordan) and anti-hermitian (Lie) subspaces as 
\begin{equation}\CLtwo = \mathcal{H}_2(\C) \oplus  \label{split2}
 \mathfrak{u}(2),
 \end{equation}
\noindent  respectively.  We will refer to equation~(\ref{split2}) as a \it Lie-Jordan splitting \rm of $\CLtwo.$  These structures are interesting in the context of $C^*$ algebras.

As described in reference~\citep{321}, the product defined as $m(a,b)=ab+ba^{\dagger}$ has
\begin{equation} \begin{array}{lll}
m(J_i, J_i)=\{J_i,J_j\} &\hspace{1cm} &\in \mathcal{H}_2(\C)\vspace{2mm}\\
m(L_i, L_j)=[L_i,L_j] &\hspace{1cm} &\in \mathfrak{u}(2)\vspace{2mm}\\
m(J_i, L_j)=\{J_i,L_j\} &\hspace{1cm} &\in \mathfrak{u}(2)\vspace{2mm}\\
m(L_i, J_j)=[L_i,J_j] &\hspace{1cm} &\in \mathcal{H}_2(\C)
\end{array}\end{equation}    
\noindent for $J_k\in\mathcal{H}_2(\C)$ and $L_k\in \mathfrak{u}(2).$  Here, $\{\hspace{.5mm}a,\hspace{.5mm}b\hspace{.5mm}\}:=ab+ba$ defines the Jordan product (up to a factor of 1/2), while $[\hspace{.5mm}a,\hspace{.5mm}b\hspace{.5mm}]:=ab-ba$ defines the Lie product.

Finally, we point out that this physically relevant matrix algebra, $M_2(\C),$ has a natural division algebraic representation.  Namely, it is known that $M_2(\C)$ is isomorphic to the complex quaternions, $\CH$, which we will introduce shortly.

\subsection{$\mathcal{H}_{16}(\C)$\label{H16}}

We would now like to extend $\mathcal{H}_2(\C)$ to a Euclidean Jordan algebra with enough capacity to accommodate all of the known local degrees of freedom of the Standard Model.  We converge on $\mathcal{H}_{16}(\C)$, the Jordan algebra of $16\times16$ hermitian matrices.  

In generalizing Section~\ref{h2c}, we may define four $16\times 16$ matrices $\alpha_1,$ $\alpha_2,$ $\alpha_3,$ $\alpha_4$ such that
\begin{equation}\label{anticom} \{  \alpha_i, \alpha_j  \} = 0, \hspace{1cm} \{  \alpha_i, \alpha_j^{\dagger}  \} = \delta_{ij}, \hspace{10mm} i,j \in \{1,2,3,4\}.
\end{equation}
\noindent   The new algebraic vacuum then becomes
\begin{equation} v:= \alpha_1\alpha_2\alpha_3\alpha_4 \alpha^{\dagger}_4 \alpha^{\dagger}_3\alpha^{\dagger}_2\alpha^{\dagger}_1.
\end{equation}
\noindent From here, a 16 $\C$-dimensional spinor may be defined as
\begin{equation}\psi := \sum_{a,b,c,d = 0}^4 c_{abcd} \hspace{.5mm}\alpha^{\dagger}_a\alpha^{\dagger}_b\alpha^{\dagger}_c\alpha^{\dagger}_dv,
\end{equation}
\noindent where $c_{abcd}\in\C$ and $\alpha_0 = \alpha_0^{\dagger} = \mathbb{I}$ represents this time the $16\times 16$ identity matrix.  A generic element of $\mathcal{H}_{16}(\C)$ may then be written as 
\begin{equation}  \sum_{\mu=1}^{256} p'_{\mu} \sigma'^{\mu} = \sum_{i=1}^{16} \psi_i \psi_i^{\dagger},
\end{equation} 
\noindent or finally,
\begin{equation}  \label{outer16}
 \sum_{\mu=1}^{256}p'_{\mu} \sigma'^{\mu} =  c_{ijklmnpq} \hspace{.5mm}\alpha_i^{\dagger}\alpha_j^{\dagger}\alpha_k^{\dagger}\alpha_l^{\dagger} \hspace{.5mm}v  \hspace{.5mm}\alpha_m\alpha_n\alpha_p\alpha_q \hspace{1mm}+\hspace{1mm} h.c.,
 \end{equation} 
\noindent for hermitian basis elements $\sigma'^{\mu}$ and  $c_{ijklmnpq}\in\C.$ Here we have an implied sum over $i,j,k,l,m,n,p,q\in\{0,1,2,3,4\}.$  

The algebra of 16$\times$16 complex matrices, $M_{16}(\C)$ coincides with the complex Clifford algebra $\CLeight,$ when thought of as a matrix algebra under multiplication $m(a,b)=ab$  for $a,b\in\CLeight$.  With regard to the hermitian conjugate, $M_{16}(\C)\simeq \CLeight$ splits into hermitian (Jordan) and anti-hermitian (Lie) subspaces as 
\begin{equation}
\CLeight = \mathcal{H}_{16}(\C) \oplus \mathfrak{u}(16), \label{split8}
 \end{equation}
 \noindent respectively.  We will refer to equation~(\ref{split8}) as a \it Lie-Jordan splitting \rm of $\CLeight.$
 
 As described in reference~\citep{321}, the product defined as $m(a,b)=ab+ba^{\dagger}$ has
\begin{equation} \begin{array}{lll}\label{multiaction}
m(J_i, J_i)=\{J_i,J_j\} &\hspace{1cm} &\in \mathcal{H}_{16}(\C)\vspace{2mm}\\
m(L_i, L_j)=[L_i,L_j] &\hspace{1cm} &\in \mathfrak{u}(16)\vspace{2mm}\\
m(J_i, L_j)=\{J_i,L_j\} &\hspace{1cm} &\in \mathfrak{u}(16)\vspace{2mm}\\
m(L_i, J_j)=[L_i,J_j] &\hspace{1cm} &\in \mathcal{H}_{16}(\C)
\end{array}\end{equation}    
\noindent for $J_k\in\mathcal{H}_{16}(\C)$ and $L_k\in \mathfrak{u}(16).$

At this point, we have introduced the Jordan algebra meant to describe the Standard Model's local degrees of freedom.  However, on its own, $\mathcal{H}_{16}(\C)$ is largely featureless.  In the next section, we will demonstrate how one can generate $\mathcal{H}_{16}(\C)$ in a way that imparts on it useful substructure needed to distinguish particle irreps.  Explicitly,  we introduce certain normed division algebras, and their multiplication algebras, although it should be noted that this article can still be understood simply in terms of   hermitian  matrices.

\section{$\mathcal{H}_{16}(\C)$ as generated by the \\division algebras\label{HLA}}

\subsection{Definition of $\A=\CHO$}

All tensor products will be over $\mathbb{R}$ throughout this article, unless otherwise stated.

As emphasized earlier, the algebra of complex quaternions, $\CH,$ is isomorphic to the algebra of 2$\times$2 complex matrices, and hence is highly relevant in theoretical physics.  We will write the usual  $\C$-basis for $\CH$ as $\{ \epsilon_0, \epsilon_1, \epsilon_2, \epsilon_3\}$.  Here, $\epsilon_0 = 1$, $\epsilon_j^2 = -1$ for $j \in \{1, 2, 3\}$, and $\epsilon_1\epsilon_2 = \epsilon_3,$ plus cyclic permutations. 
These last relations may be captured more succinctly as 
\begin{equation}\epsilon_a\epsilon_b = -\delta_{ab}+\varepsilon_{abc}\hspace{0.5mm}\epsilon_c
\end{equation} 
\noindent for $a,b,c\in \{1,2,3\}$, where $ \varepsilon_{abc}$ is the usual totally anti-symmetric tensor with $ \varepsilon_{123}=1$.  We will take $i$ to denote the complex imaginary unit as usual.

It is less well-known that the complex octonions, $\CO,$ are also relevant for theoretical physics.  See for example G\"{u}naydin and G\"{u}rsey's early work on quarks~\citep{GGquarks}.  We will write the usual $\C$-basis for the complex octonions, $\C\otimes\mathbb{O},$ as $\{e_0, e_1, \dots, e_7\}$.  Here, $e_0 = 1$ and 
\begin{equation}e_ie_j = -\delta_{ij}+f_{ijk}\hspace{0.5mm}e_k,  
\end{equation}
\noindent for $i,j,k\in \{1,2,\dots 7\}$.  Analogously, $ f_{ijk}$ is a totally anti-symmetric tensor with $f_{ijk}=1$ when $ijk\in \{124, 235, 346, 457, 561, 672, 713\}$.   The remaining values of $f_{ijk}$ are determined by anti-symmetry, and vanish otherwise.

We can now join $\CH$ and $\CO$ together in a tensor product over $\C$.  This gives $\A :=(\CH)\otimes_{\C}(\CO) = \CHO$.  Multiplication of elements in $\A$ is defined by setting $(x_1 \otimes_{\C} y_1)(x_2 \otimes_{\C} y_2) = x_1x_2 \otimes_{\C} y_1y_2$ for all $ x_1, x_2 \in \CH$, and $y_1, y_2 \in \CO$. A $\C$-basis for $\A$ is given by $\{\epsilon_j \otimes e_k \mid j \in \{0, 1, 2, 3\}, k \in \{0, 1, \dots, 7\}\}$. Readers can then confirm  that $\A$ is a  32\hspace{0.5mm}$\C$-dimensional non-commutative, non-associative algebra. 

For convenience, we will write arbitrary elements of $\A$ as $\sum_{j, k} c_{jk}\hspace{.5mm}\epsilon_je_k$ for $c_{jk}\in\C$, now omitting the $\otimes$ symbol. We see that $\epsilon_je_k = e_k\epsilon_j$ for all $j, k$.   As an example calculation, consider $a,b\in\A$ with  $a=4 \epsilon_1 e_2$ and $b=\left(5+i\right) \epsilon_2 e_4$.  Then $ab = \left(20+4i\right)\epsilon_3e_1.$

Recently  it was shown how $\A=\CHO$ can encode a full generation of Standard Model particles,~\citep{fh1}, \citep{dasb}.

\subsection{Definition of $\A$'s  multiplication algebras \label{multalg}}

With this said, our ultimate goal is to encode not only a single generation, but rather, the Standard Model's three generations, together with its gauge and Higgs bosons.  How might one leverage $\CHO$ to somehow accommodate so many more degrees of freedom?  

Inspired by nature's genetic code, \citep{Fur2023}, let us consider \it sequences \rm of $\CHO$ elements left-multiplying themselves.

Recall that $\C l (8)$ is the 256-dimensional unital associative algebra over $\C$ generated by  $\gamma_1, \dots, \gamma_8,$ satisfying the relations 
\begin{equation}\{ \gamma_j, \gamma_k\} = 2\delta_{jk}
\end{equation}
\noindent for all $j, k$. Given $x \in \A$, we write $L_x$ for the complex-linear map $\A \to \A$ given by 
\begin{equation}L_x(y) := xy
\end{equation} 
\noindent for all $y \in \A$. 

We define $\A$'s multiplication algebra, $L_{\A},$ to be the subalgebra of $End_{\C}(\A)$ generated by $\{L_x \mid x \in \A\}$.  It is straightforward to show that $L_{\A}$ is isomorphic to $\C l (8)$. To be explicit, one may take the eight elements
\begin{equation}\label{gammas}
\{\hspace{.5mm}L_{ie_j}, \hspace{.5mm}L_{\epsilon_1e_7}, \hspace{.5mm}L_{\epsilon_2 e_7}\hspace{.5mm}\}
\end{equation}  
\noindent for $j \in \{1, 2, \dots, 6\}$ as a generating set.  As an example calculation in $L_{\A}$, consider the multiplication of two elements $A,B\in End(\A)$ with $A=4 L_{\epsilon_1e_2}$ and $B=\left(5+i\right) L_{\epsilon_2 e_4}$.  Then $AB:=A\circ B = \left(20+4i\right)L_{\epsilon_1e_2}\circ L_{\epsilon_2 e_4} = \left(20+4i\right)L_{\epsilon_3e_2}\circ L_{e_4}\neq  \left(20+4i\right)L_{\epsilon_3e_1}.$

For later use, we may define $\A$'s right multiplication algebra, $R_{\A},$ to be the subalgebra of $End_{\C}(\A)$ generated by $\{R_x \mid x \in \A\},$ where $R_x(y) := yx.$  Finally, we define $\A$'s full multiplication algebra, $M_{\A},$ to be the subalgebra of $End_{\C}(\A)$ generated by both $\{L_x \mid x \in \A\}$ and $\{R_{x'} \mid x' \in \A\}.$  It can be confirmed that $M_{\A}\simeq End_{\C}(\A).$

Given $x \in L_{\A}$, we write $x^*$ for the complex conjugate of $x$ that maps $i\mapsto -i.$  We also define an anti-automorphism $L_{\A} \to L_{\A},$ denoted $x \mapsto x^\dagger,$ by $i \mapsto -i$, $L_{\epsilon_j} \mapsto -L_{\epsilon_j}$ for $j \in \{1, 2, 3\}$, $L_{e_k} \mapsto -L_{e_k}$ for $k \in \{1, \dots, 7\}$, with $(xy)^\dagger = y^\dagger x^\dagger$ for all $x, y \in L_{\A}$. We call $x^\dagger$ the \it hermitian conjugate \rm of $x$.  

In this article, we will be most interested in the hermitian subset of $L_{\A},$ that is, $\mathcal{H}_{L_\A}:=\{h\in L_{\A} \mid h=h^{\dagger}\}$.  It is possible to show that $\mathcal{H}_{L_\A}$ forms an algebra under the Jordan product $m(a,b)=ab+ba = ab+ba^{\dagger}$ for $a,b\in\mathcal{H}_{L_\A},$  and that this algebra is isomorphic to the  16$\times$16 hermitian matrices, $\mathcal{H}_{16}(\C).$

We will find that $\mathcal{H}_{L_{\CH}}$ encodes the representations of spacetime symmetries, while $\mathcal{H}_{L_{\CO}}$ encodes the representations of internal symmetries.  This identification allows for easy compliance with the Coleman-Mandula theorem.

\section{Early three generation results\label{early}}

\subsection{Three generation $\mathfrak{su}(3)_C$ structure}

Finding mathematical objects with natural three-generation structure has, over the years, persistently proven a challenge.  With this said, some preliminary QCD results were found in~\citep{Gen}, \citep{thesis}, \citep{321}.

The algebra in which this $\mathfrak{su}(3)_C$ structure was identified is the left multiplication algebra of the complex octonions, $L_{\CO}.$ Readers can confirm that $L_{\CO}$ is isomorphic to the complex Clifford algebra $\CLsix,$ in addition to being isomorphic to $End_{\C}(\CO)$.  It may be generated by the set $\{L_{e_j}\}$ for $j\in\{1\dots 6\}.$  In this case, it can be shown that the linear map $L_{e_7}$ coincides with $L_{e_1}L_{e_2}L_{e_3}L_{e_4}L_{e_5}L_{e_6},$ known as the volume element of the Clifford algebra.

Now incidentally, this same octonionic imaginary unit, $e_7,$ provides us with another Clifford volume element, $R_{e_7},$ when $\CLsix$ is generated by the set $\{R_{e_j}\}$ for $j\in\{1\dots 6\}.$  That is, $R_{e_1}R_{e_2}R_{e_3}R_{e_4}R_{e_5}R_{e_6}=-R_{e_7}.$  Note that $R_{e_7}\neq L_{e_7}.$  The  elements $L_{ie_7}$ and $R_{ie_7}$ may be thought of as generalizations of the chirality operator $\gamma^5:=i\gamma^0\gamma^1\gamma^2\gamma^3$ in the Dirac algebra.  As we will see, \it they supply the division algebraic substructure that defines the $\mathfrak{su}(3)_C$ particle representations. \rm

It should be noted that $\{L_{e_j}\}$ and $\{R_{e_j}\}$ for $j\in\{1\dots 6\}$ are not the only generators capable of producing $L_{e_7}$ and $R_{e_7}$ as volume elements.  In keeping with the well-known table of Clifford matrix algebras,  e.g. \citep{lounesto} or \citep{thesis} p. 20, there are three distinct ways to generate $L_{\mathbb{O}}\simeq M_8(\mathbb{R}).$  They correspond to Clifford algebras $Cl(0,6),$ $Cl(3,3),$ and $Cl(4,2).$  Not only can $Cl(0,6)$ lead to Clifford volume elements $L_{e_7}$ and $R_{e_7}$, as described above, but so can $Cl(4,2).$  Consider, for example $Cl(4,2)$ as generated by $\{\hspace{.5mm} L_{e_2},\hspace{.5mm} L_{e_6}, \hspace{.5mm}L_{e_1}L_{e_3}L_{e_4}, \hspace{.5mm}L_{e_1}L_{e_3}L_{e_5},\hspace{.5mm} L_{e_1}L_{e_4}L_{e_5}, \hspace{.5mm}L_{e_3}L_{e_4}L_{e_5}\hspace{.5mm} \}.$  Of course, this signature is of special importance due to its connection to conformal symmetry in 3+1 D.

These Clifford volume elements allow us to define a set of idempotents, 
\begin{equation}\begin{array}{lll}
s:=\frac{1}{2}\left(1+iL_{e_7}\right) &\hspace{2mm}& s^*:=\frac{1}{2}\left(1-iL_{e_7}\right)\vspace{2mm}\\
S:=\frac{1}{2}\left(1+iR_{e_7}\right) &\hspace{2mm}& S^*:=\frac{1}{2}\left(1-iR_{e_7}\right),
\end{array}\end{equation}
\noindent that break $L_{\CO}$ into blocks.  More precisely, the set $\{\hspace{.5mm}sS, \hspace{.5mm}s^*S, \hspace{.5mm}sS^*, \hspace{.5mm}s^*S^* \}$ facilitates what is known as a \it Peirce decomposition, \rm which we will define more carefully in the context of $\mathcal{H}_{L_{\A}}.$

\begin{figure}[h]
\begin{center}
\includegraphics[width=9cm]{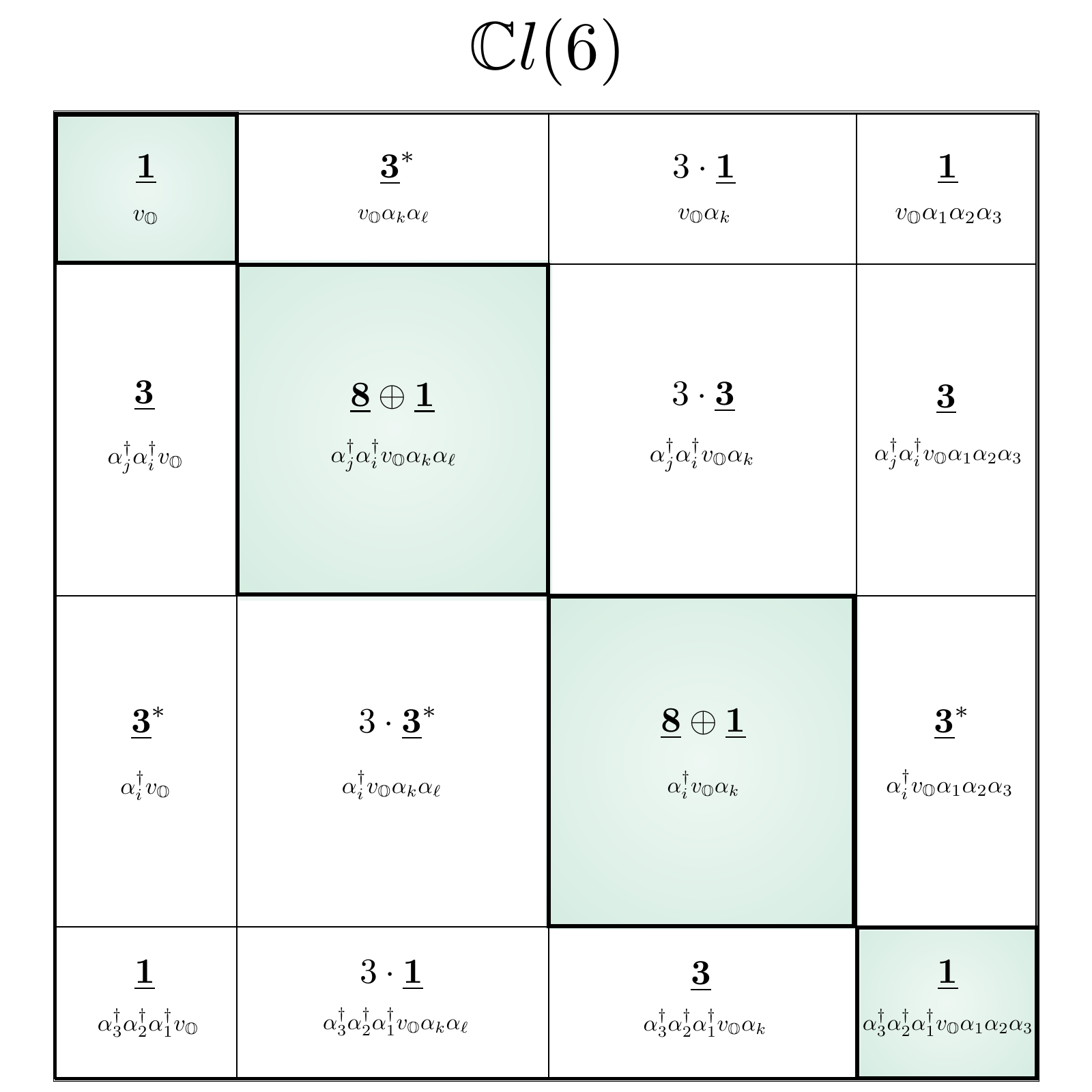}
\caption{\label{CL6}  Decomposition of  $L_{\CO}\simeq \CLsix$ into  $\mathfrak{su}(3)_C$ representations familiar from gluons, and three generations of quarks and leptons.}
\end{center}\end{figure}

Setting 
\begin{equation}\begin{array}{lll}\label{octladders}
\alpha_1:=\frac{1}{2}\left( -L_{e_5} +iL_{e_4}\right)&\hspace{2mm}&\alpha_1^{\dagger}:=\frac{1}{2}\left( L_{e_5} +iL_{e_4}\right)\vspace{2mm}\\
\alpha_2:=\frac{1}{2}\left( -L_{e_3} +iL_{e_1}\right)&&\alpha_2^{\dagger}:=\frac{1}{2}\left( L_{e_3} +iL_{e_1}\right)\vspace{2mm}\\
\alpha_3:=\frac{1}{2}\left( -L_{e_6} +iL_{e_2}\right)&&\alpha_3^{\dagger}:=\frac{1}{2}\left( L_{e_6} +iL_{e_2}\right),
\end{array}\end{equation}
\noindent and
\begin{equation}
v_{\mathbb{O}}:=\alpha_1\alpha_2\alpha_3\alpha_3^{\dagger}\alpha_2^{\dagger}\alpha_1^{\dagger},
\end{equation}
\noindent we may then identify $\mathfrak{su}(3)_C$ generators as
\begin{equation}\begin{array}{l}\label{liealg}
i\Lambda_1:= -i\alpha_2^{\dagger}\alpha_3^{\dagger}\hspace{.5mm}v_{\mathbb{O}}\hspace{.5mm}\alpha_3\alpha_1    -i\alpha_1^{\dagger}\alpha_3^{\dagger}\hspace{.5mm}v_{\mathbb{O}}\hspace{.5mm}\alpha_3\alpha_2,         \vspace{2mm}          

\\    i\Lambda_2:=-\alpha_2^{\dagger}\alpha_3^{\dagger} \hspace{.5mm}  v_{\mathbb{O}} \hspace{.5mm}\alpha_3\alpha_1    +\alpha_1^{\dagger}\alpha_3^{\dagger} \hspace{.5mm}  v_{\mathbb{O}} \hspace{.5mm}\alpha_3\alpha_2,         \vspace{2mm} 

\\  i\Lambda_3:=-i\alpha_2^{\dagger}\alpha_3^{\dagger} \hspace{.5mm}  v_{\mathbb{O}} \hspace{.5mm}\alpha_3\alpha_2    -i\alpha_1^{\dagger}\alpha_3^{\dagger} \hspace{.5mm}  v_{\mathbb{O}} \hspace{.5mm}\alpha_3\alpha_1,         \vspace{2mm} 
\\      i\Lambda_4:=-i\alpha_1^{\dagger}\alpha_2^{\dagger} \hspace{.5mm}  v_{\mathbb{O}} \hspace{.5mm}\alpha_2\alpha_3    -i\alpha_3^{\dagger}\alpha_2^{\dagger} \hspace{.5mm}  v_{\mathbb{O}} \hspace{.5mm}\alpha_2\alpha_1,         \vspace{2mm} 
 
\\ i\Lambda_5:=\alpha_1^{\dagger}\alpha_2^{\dagger} \hspace{.5mm}  v_{\mathbb{O}} \hspace{.5mm}\alpha_2\alpha_3    -\alpha_3^{\dagger}\alpha_2^{\dagger} \hspace{.5mm}  v_{\mathbb{O}} \hspace{.5mm}\alpha_2\alpha_1,         \vspace{2mm} 
  
\\   i\Lambda_6:=-i\alpha_3^{\dagger}\alpha_1^{\dagger} \hspace{.5mm}  v_{\mathbb{O}} \hspace{.5mm}\alpha_1\alpha_2    -i\alpha_2^{\dagger}\alpha_1^{\dagger} \hspace{.5mm}  v_{\mathbb{O}} \hspace{.5mm}\alpha_1\alpha_3,         \vspace{2mm} 

\\ i\Lambda_7:=-\alpha_3^{\dagger}\alpha_1^{\dagger} \hspace{.5mm}  v_{\mathbb{O}} \hspace{.5mm}\alpha_1\alpha_2    +\alpha_2^{\dagger}\alpha_1^{\dagger} \hspace{.5mm}  v_{\mathbb{O}} \hspace{.5mm}\alpha_1\alpha_3,         \vspace{2mm}

\\ i\Lambda_8:=  \frac{i}{\sqrt{3}}\left(\alpha_1^{\dagger}\alpha_3^{\dagger} \hspace{.5mm}  v_{\mathbb{O}} \hspace{.5mm}\alpha_3\alpha_1 
+ \alpha_2^{\dagger}\alpha_3^{\dagger} \hspace{.5mm}  v_{\mathbb{O}} \hspace{.5mm}\alpha_3\alpha_2 - 2\alpha_1^{\dagger}\alpha_2^{\dagger} \hspace{.5mm}  v_{\mathbb{O}} \hspace{.5mm}\alpha_2\alpha_1\right).
\vspace{2mm}
\end{array}\end{equation}
\noindent  These generators commute with $L_{e_7},$ $R_{e_7},$ and reside within the $sS^* \hspace{.5mm}L_{\CO}\hspace{.5mm}sS^*$  subspace.  That is, they are given by $i\lambda_jsS^*,$ where the $i\lambda_j$ generate $\mathfrak{su}(3) \subset \mathfrak{g}_2=\mathfrak{der}(\mathbb{O}),$ as described in~\citep{thesis}.  Under a simple action
\begin{equation} 
\left[\hspace{.5mm}\hspace{.5mm}r_j i\Lambda_j,\hspace{.5mm} sL_{\CO} \hspace{.5mm}\hspace{.5mm}\right] +  \left[\hspace{.5mm}\hspace{.5mm}-r_j i\Lambda_j^*, \hspace{.5mm}s^*L_{\CO}\hspace{.5mm}\hspace{.5mm} \right],
\end{equation}
\noindent the algebra $L_{\CO}\simeq \CLsix$ decomposes as
\begin{equation}\begin{array}{rllllllllll} \label{COdecomp}
 (\mathbf{\underline{8}})&\oplus&(3\times \mathbf{\underline{3}})&\oplus&(3\times \mathbf{\underline{3}})&\oplus&(3\times \mathbf{\underline{1}})&\oplus&(3\times \mathbf{\underline{1}})&\oplus\vspace{2mm}\\

(\mathbf{\underline{8}})&\oplus&(3\times \mathbf{\underline{3}}^*)&\oplus&(3\times \mathbf{\underline{3}}^*)&\oplus&(3\times \mathbf{\underline{1}})&\oplus&(3\times \mathbf{\underline{1}}).&\vspace{2mm}
\end{array}\end{equation}
\noindent Said another way, a simple $\mathfrak{su}(3)_C$ action fragments $\CLsix$ into QCD representations familiar from \it gluons and three generations of fermions. \rm Please see Figure~(\ref{CL6}). 

\subsection{Spacetime structure}

Of course, $\mathfrak{su}(3)_C$ representations are not sufficient to claim a full description of the Standard Model's particle content.  We  also require electroweak and spacetime representations.

In order to include these spacetime symmetries, it was proposed in~\citep{Gen} that $L_{\CO}$ be extended to 
\begin{equation}L_{\CHO}=L_{\A}\simeq\CLeight, 
\end{equation}
\noindent thereby finally incorporating the quaternions.   The idea, maintained here, is that \it quaternions should encode spacetime symmetries, while octonions should encode internal gauge symmetries.\rm

For subsequent three-generation $\CLeight$ proposals, please see articles by Gillard and Gresnigt,~\citep{niels_sedenion},  Gording and Schmidt-May,~\citep{gsm}.

Building from the $L_{\A}\simeq\CLeight$ generators of expression~(\ref{gammas}), let us finally establish a full set of non-trivial ladder operators, $\{\alpha_1, \alpha_2, \alpha_3, \alpha_4, \alpha_1^{\dagger}, \alpha_2^{\dagger},\alpha_3^{\dagger},\alpha_4^{\dagger} \},$ as proposed in Section~\ref{H16}.  We will take the set $\{\alpha_1, \alpha_2, \alpha_3,  \alpha_1^{\dagger}, \alpha_2^{\dagger},\alpha_3^{\dagger} \}$  to be given by equations~(\ref{octladders}).   Furthermore, we define
\begin{equation}\begin{array}{l}
\alpha_4:=\beta_4:=\frac{1}{2}\left( -L_{\epsilon_2}+iL_{\epsilon_1}\right)L_{ie_7}, \vspace{2mm}\\
\alpha_4^{\dagger}:=\beta_4^{\dagger}:=\frac{1}{2}\left( L_{\epsilon_2}+iL_{\epsilon_1}\right)L_{ie_7}, 
\end{array}\end{equation}
\noindent where we have named these final ladder operators as $\beta_4$ and $\beta_4^{\dagger}$ so as to distinguish these (mostly) quaternionic operators from their octonionic counterparts, $\{\alpha_1,  \alpha_2,  \alpha_3, \alpha_1^{\dagger},\alpha_2^{\dagger},\alpha_3^{\dagger} \}.$  We define the (purely) quaternionic vacuum as 
\begin{equation}
v_{\mathbb{H}}:=\beta_4\beta_4^{\dagger}.
\end{equation}

\subsection{Towards electroweak structure \label{EW}}

 It should be noted that the decomposition~(\ref{COdecomp}) contains an extra eight-dimensional adjoint representation as compared to the Standard Model.  Furthermore we have yet to describe the electroweak sector.  Could the additional $\mathbf{\underline{8}}$ somehow be broken down so as to introduce familiar electroweak structure?  Indeed, there is a well-known decomposition, 
\begin{equation}\begin{array}{lll} \label{bd}
\mathfrak{su}(3)&&\hspace{3mm}\mathfrak{su}(2)\vspace{2mm}\\
\underline{\mathbf{8}}&\mapsto& \hspace{3mm}\underline{\mathbf{3}}\oplus \underline{\mathbf{2}} \oplus \underline{\mathbf{2}} \oplus \underline{\mathbf{1}},
\end{array}\end{equation}
\noindent first introduced to this author by B. Gording.  In fact, it is most likely that this idea was incorporated in a $\CLsix$ model proposed by Gording and Schmidt-May in~\citep{gsm}.

As will be later demonstrated,  this decomposition can be effected by making use of another admissible volume element of $L_{\CO}.$  That is, generating $L_{\CO}\simeq \CLsix$ with the set $\{L_{e_1}, L_{e_3}, L_{e_4}, L_{e_2}L_{e_5}L_{e_6},L_{e_5}L_{e_6}L_{e_7}, L_{e_2}L_{e_5}L_{e_7}\}$ produces a Clifford volume element $L_{e_2}L_{e_6}L_{e_7}.$  This description of $\CLsix$ clearly arises as the complexification $Cl(3,3)$.  The three octonionic imaginary units $\{e_2, e_6, e_7\}$ define a quaternionic subalgebra in $\mathbb{O}$.  More minimalistically, a quaternionic subalgebra may be specified by choosing a second octonionic imaginary unit, say $e_6,$ orthogonal to a first, say $e_7,$ see  \citep{mosaic}, \citep{FR2022}.

A number of authors have advocated for $Cl(3,3)$ in various ways, including for its chiral properties.  Please see Chester, Rios, Marrani~\citep{Che2020}, Wilson \citep{wilson}, and  Vaibhav, Singh \citep{singh_v}, and most recently, Betancourt,~\citep{bet}.  Also highly relevant is a one-generation non-commutative geometry model  proposed by Barrett, \citep{ncgcliff}.

One can construct a number of proposals to explain the origin of this quaternionic substructure.  For example, one may adapt the model by considering $\mathbb{O}\oplus \mathbb{O}$ instead of $\CO$ as the algebra from which one builds multiplication algebras.  From here, one may consider preserved complex structures $L_{e_7}$ and $R_{e_7}$ on the first copy of $\mathbb{O},$ and a preserved quaternionic structure on the second.

Alternatively, one may wish to consider the multiplication algebra of $\OHCR,$ or further, that of $\mathbb{S},$ with the nested embeddings $\R \subset \C\subset \mathbb{H}\subset \mathbb{O}$ in $\mathbb{S},$ as detailed in section~\ref{plus}.

\section{Standard Model representations inside $\mathcal{H}_{16}(\C)$\label{assemble}}

Finally we are ready to upgrade $L_{\CO}\simeq \CLsix$ to the full  $L_{\CHO}\simeq\CLeight$ algebra, and to take its hermitian part, $\mathcal{H}_{L_{\A}} \simeq \mathcal{H}_{16}(\C)$.  We will see that the Clifford volume elements $\{L_{e_7},$  $R_{e_7},$  $L_{e_2}L_{e_6}L_{e_7}\}$ \it induce a splitting of $\mathcal{H}_{L_{\A}}$ into a set of vector spaces that mirror, to a large extent, familiar Standard Model irreps. \rm More formally, this splitting is known as a \it Peirce decomposition. \rm See Figure~(\ref{map}) for a preview.

\subsection{Peirce decomposition \label{PD}}

Given a set $ \{ s_1, \dots, s_n\}$ of mutually orthogonal idempotents in $\mathcal{H}_{16}(\C)$ satisfying $\sum_{i = 1}^n s_i = 1$, one may form the \emph{Peirce decomposition} of $\mathcal{H}_{16}(\C)$ as
\begin{equation}
\mathcal{H}_{16}(\C) = \underset{1\leq j \leq k \leq n}\bigoplus \mathcal{H}_{jk}
\end{equation}
with respect to $ \{ s_1, \dots, s_n\}$. The subspaces $\mathcal{H}_{jk}$ are defined as follows: we have 
\begin{equation}\mathcal{H}_{jj} = s_j\mathcal{H}_{16}(\C)s_j \label{diag} \end{equation}
\noindent for all $j$. When $j \neq k$, we have
\begin{equation}\label{offdiag}
\mathcal{H}_{jk} = \{ s_j x s_k + s_k x s_j \mid x \in \mathcal{H}_{16}(\C) \}.
\end{equation}

In the decomposition of $L_{\CO}$ into $\mathfrak{su}(3)$ representations of Figure~\ref{CL6}, it is clear that the set of  idempotents 
\begin{equation}\mathcal{S}_4:=\{\hspace{.5mm} sS, \hspace{.5mm}sS^*, \hspace{.5mm}s^* S, \hspace{.5mm}s^* S^*\}\end{equation}
\noindent induced a Peirce decomposition of $L_{\CO}.$

Defining now
\begin{equation}\begin{array}{l}S_a:=\frac{1}{2}\left(1+L_{e_2}L_{e_6}L_{e_7}\right),\hspace{3mm}
 S_{\bar{a}}:=\frac{1}{2}\left(1-L_{e_2}L_{e_6}L_{e_7}\right),\vspace{1mm}
\end{array}\end{equation}
\noindent we will consider the Peirce decomposition of $\mathcal{H}_{16}(\C)$ with respect to the set 
\begin{equation}\label{S5}
\mathcal{S}_5:=\{\hspace{.5mm} sS,\hspace{.5mm} sS^*, \hspace{.5mm}s^*S\hspace{.5mm}S_a, \hspace{.5mm}s^*S\hspace{.5mm}S_{\bar{a}}, \hspace{.5mm}s^* S^*S_{\bar{a}}\hspace{.5mm} \}.
\end{equation} 
\noindent We  label these idempotents as $s_1:=sS,$ $s_2:=sS^*,$ $s_3:=s^*S\hspace{.5mm}S_a,$ $s_4:=s^*S\hspace{.5mm}S_{\bar{a}},$ $s_5:=s^*S^*S_{\bar{a}}=s^*S^*.$  It can be confirmed that this is a set of mutually orthogonal idempotents that sum to $1$.  

Readers will find later in this article that the change from  $\mathcal{S}_4$ to $\mathcal{S}_5$ indeed induces the electroweak decomposition of expression~(\ref{bd}).  This change in Peirce sets induces a further partitioning of the Peirce blocks in Figure~(\ref{CL6}).  We will make those new partition lines known by depicting them in blue.  Relevant figures where these blue partition lines may be seen include Figures~(\ref{outeroff}),  (\ref{diagfig}), (\ref{inneroff}), (\ref{map}). 

Beyond isolating quaternionic subspaces within the octonions, we point out that $S_a$ and $S_{\bar{a}}$ idempotents help encode a certain $i\hspace{.5mm}\mathfrak{su}(5)$ algebra within the $\mathcal{H}_{22},\mathcal{H}_{23}, \mathcal{H}_{33}$ blocks of Figure~\ref{map}.   See~\citep{AGUTS} and \citep{roadI}.   (It should be noted that Kaushal Kumar participated in a calculation of these idempotents in the context of a related non-commutative geometry project.)

A natural question to ask at this point is:  What motivates these particular $\mathcal{S}_4$ and $\mathcal{S}_5$ Peirce sets?  At the moment we will mention that  $\mathcal{S}_4$ derives from identifying certain complex structures within the octonionic sector, while $S_a$ and $S_{\bar{a}}$ derive from identifying a certain quaternionic structure.  Please see Sections~\ref{EW} and~\ref{plus} for relevant proposals.

\subsection{Symmetries}

With a division algebraic Peirce decomposition fixed, we would now like to determine:  which symmetries respect it?  As mentioned earlier, the set $L_{\A}\simeq \CLeight$ may be viewed as 
\begin{equation}\CLeight = \mathcal{H}_{16}(\C) \oplus \mathfrak{u}(16),
\end{equation}
\noindent whereby  $\mathfrak{u}(16)$ acts on $\mathcal{H}_{16}(\C)$ via a commutator, per equations~(\ref{multiaction}) and reference \citep{321}.

The internal symmetries of our system will be represented by those elements $r_au_a\in\mathfrak{u}(16)$ obeying
\begin{equation}
ir_au_a\hspace{1mm}\in \hspace{1mm}s_j\hspace{.5mm} \mathcal{H}_{L_{\CO}}\hspace{.5mm}s_j
\end{equation}
\noindent for idempotent $s_j\in \mathcal{S}_5,$ and $r_a\in\mathbb{R}.$ On the other hand, spatial symmetries of our system obey
\begin{equation}
ir_au_a\hspace{1mm}\in\hspace{1mm} s_j\hspace{.5mm} \mathcal{H}_{L_{\CH}}\hspace{.5mm}s_j.
\end{equation}
\noindent This natural octonion-quaternion splitting allows for easy compliance with the Coleman-Mandula theorem.  

Clearly, these symmetries commute with all idempotents in $\mathcal{S}_5.$  Readers may verify that octonionic (internal) symmetries then take the form
\begin{equation}\begin{array}{lll}\label{osym}
is_1 \hspace{.5mm}\mathcal{H}_{L_{\CO}} \hspace{.5mm}s_1 &\simeq & \mathfrak{u}(1),\vspace{2mm}\\
is_2 \hspace{.5mm}\mathcal{H}_{L_{\CO}} \hspace{.5mm}s_2 &\simeq & \mathfrak{u}(3),\vspace{2mm}\\
is_3 \hspace{.5mm}\mathcal{H}_{L_{\CO}}\hspace{.5mm} s_3 &\simeq & \mathfrak{u}(2),\vspace{2mm}\\
is_4\hspace{.5mm} \mathcal{H}_{L_{\CO}}\hspace{.5mm} s_4 &\simeq & \mathfrak{u}(1),\vspace{2mm}\\
is_5 \hspace{.5mm}\mathcal{H}_{L_{\CO}} \hspace{.5mm}s_5 &\simeq & \mathfrak{u}(1).
\end{array}\end{equation}

 We identify the Standard Model's $\mathfrak{su}(3)_C$ within $is_2 \hspace{.5mm}\mathcal{H}_{L_{\CO}} \hspace{.5mm}s_2,$ with generators as in equation~(\ref{liealg}).  

We identify the Standard Model's  $\mathfrak{su}(2)_L$ within $is_3 \hspace{.5mm}\mathcal{H}_{L_{\CO}} \hspace{.5mm}s_3,$ with generators
\begin{equation}\begin{array}{lll}
i\tau_1:=i(\alpha_1^{\dagger}v_{\mathbb{O}}\alpha_2 + \alpha_2^{\dagger}v_{\mathbb{O}}\alpha_1)&
i\tau_2:=\alpha_1^{\dagger}v_{\mathbb{O}}\alpha_2 - \alpha_2^{\dagger}v_{\mathbb{O}}\alpha_1 \vspace{2mm}\\
i\tau_3:=i(\alpha_1^{\dagger}v_{\mathbb{O}}\alpha_1  -\alpha_2^{\dagger}v_{\mathbb{O}}\alpha_2).
\end{array}\end{equation}
  
Using the conventions of~\citep{bm}, we identify weak hypercharge $\mathfrak{u}(1)_Y$ within $is_1 \hspace{.5mm}\mathcal{H}_{L_{\CO}}\hspace{.5mm} s_1 \hspace{.5mm}\oplus\hspace{.5mm} is_2\hspace{.5mm} \mathcal{H}_{L_{\CO}}\hspace{.5mm} s_2 \hspace{.5mm}\oplus\hspace{.5mm} is_4 \hspace{.5mm}\mathcal{H}_{L_{\CO}} \hspace{.5mm}s_4 \hspace{.5mm}\oplus \hspace{.5mm}is_5 \hspace{.5mm}\mathcal{H}_{L_{\CO}} \hspace{.5mm}s_5,$ with generator
\begin{equation}\begin{array}{lll}
iY&:=&\frac{i}{2}\left(  sS + \frac{1}{3} sS^* + s^*S\hspace{.5mm}S_{\bar{a}}-s^*S^*   \right),\vspace{2mm}\\
&=& \frac{i}{2}\left(  s_1 + \frac{1}{3} s_2 + s_4-s_5   \right).
\end{array}\end{equation}

Apart from including the Standard Model's internal gauge symmetries, we find that compatibility with the Peirce decomposition \it precludes those transitions leading to proton decay. \rm

On the other hand, we find that non-redundant quaternionic (spatial) symmetries take the form
\begin{equation}\begin{array}{lll}\label{5rot}
is_j \hspace{.5mm}\mathcal{H}_{L_{\CH}} \hspace{.5mm}s_j &\simeq &  \mathfrak{su}(2)=\mathfrak{so}(3) 
\end{array}\end{equation}
\noindent for all $j.$  (By `non-redundant', we mean to say that the expected quaternionic $\mathfrak{u}(2)$ can be written as $\mathfrak{u}(2)=\mathfrak{su}(2)\oplus \mathfrak{u}(1) = \mathfrak{so}(3)\oplus \mathfrak{u}(1) $.  However, this $\mathfrak{u}(1)$ is only a phase, already accounted for in the octonionic symmetries of equations~(\ref{osym}).)  There is much to say with regards to these quaternionic symmetries.  A dedicated discussion can be found in Section~\ref{disc}.

Anomaly cancellation is to be considered once a dynamical model is established.

\subsection{Weyl Super operators}

The utility of the  Peirce decomposition above is that it molds  $\mathcal{H}_{16}(\C)$ into a space of  \it Weyl super operators.  \rm  By `Weyl super operators', we mean elements of a superalgebra that generalize the familiar operator $p_{\mu}\sigma^{\mu}$ from the momentum space Weyl equation.

Explicitly, the degrees of freedom of Weyl operators $p_{\mu}\sigma^{\mu}$ and gauge bosons reside in  \emph{diagonal} $\mathcal{H}_{jj}$ blocks.  Together they form a momentum-space representation of the covariant derivative.  

On the other hand, the complex degrees of freedom of three generations are represented in the \emph{off-diagonal} subspaces, $\mathcal{H}_{jk}$ for $j\neq k$.   For fermions, particle-anti-particle partners are mapped one to the other via hermitian conjugation.   

The algebraic feature that distinguishes gauge bosons from  three generations of fermions is idempotent structure.  That is,  gauge bosons are distinguished by symmetric idempotents, as in equation (\ref{diag}), whereas three generations have non-symmetric idempotents, as in equation (\ref{offdiag}).  Given that this model is intended to be pre-geometric, with no spacetime a priori, the usual spin statistics will need to be reconstructed carefully.  This task is outside of the scope of this article.

For another research programme wherein a related superconnection is central, please see Alvarez, Delage, Valenzuela, and Zanelli,~\citep{z_ped}, \citep{z_chiral}.

In the material to follow,  the Standard Model particle representations are mapped to specific Peirce subspaces of $\mathcal{H}_{L_\A}\simeq\mathcal{H}_{16}(\C).$  There, the objects $\alpha_i^{\dagger}\alpha_j^{\dagger}\alpha_k^{\dagger}\alpha_l^{\dagger} \hspace{.5mm}v  \hspace{.5mm}\alpha_m\alpha_n\alpha_p\alpha_q$ may be viewed as basis vectors, while the $\color{gray} e_L^{\uparrow}, \nu_{eL}^{\uparrow}, \dots$ written in grey, are coefficients providing particle identification.

\newpage

\subsubsection{Outer off-diagonal blocks  \\Two generations  \hspace{8cm} $64\hspace{.5mm}\mathbb{R}+64\hspace{.5mm}\mathbb{R}$\label{outer}}

\begin{figure}[h]
\begin{center}
\includegraphics[width=7cm]{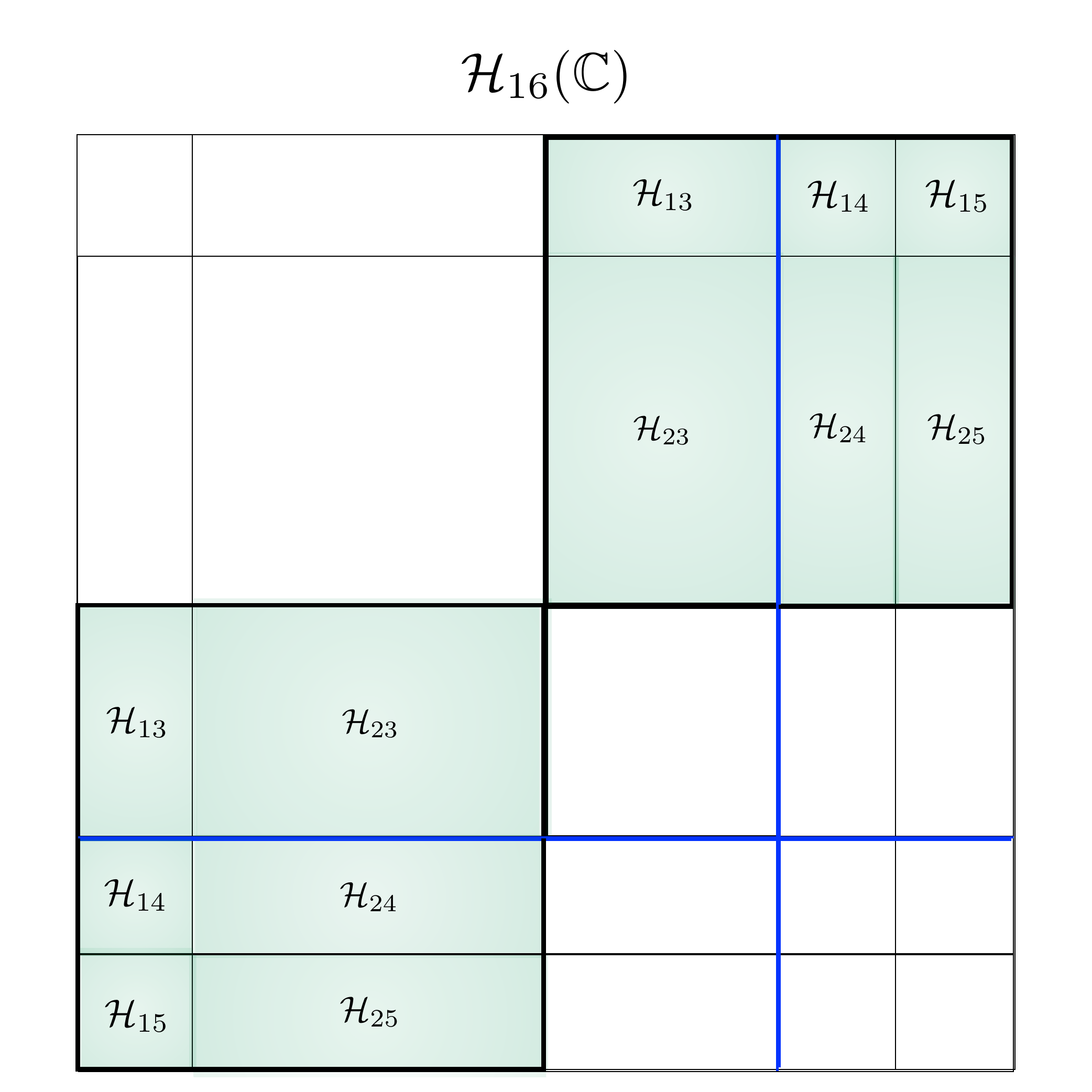}
\caption{\label{outeroff}  Two generations of off-shell fermion representations in the outer off-diagonal blocks, including sterile neutrinos. }
\end{center}\end{figure}

Applying symmetry generators as $\left[\hspace{.5mm}r_ju_j, \hspace{.5mm}  \mathcal{H}_{16}(\C)\hspace{.5mm} \right],$ we identify two generations in the outer off-diagonal blocks.  Please see Figure~(\ref{outeroff}).  This includes right-handed neutrinos.

One fundamental difference between the representations demonstrated here and those of the Standard Model lie in their behaviour under spatial symmetries.  As will be discussed in Section~\ref{disc}, the fermions in this model, $s_i \Psi s_j$,  are interpreted as extended objects within the Jordan algebra.    That is, $s_i$ and $s_j$ for $i\neq j$ may be viewed as two distinct sites.  These sites are acted upon by two independent $\mathfrak{su}(2)=\mathfrak{so}(3)$ spatial symmetries, given by equation~(\ref{5rot}).  

In the standard description of Weyl spinors, \citep{TASI}, right-handed fermions are acted upon from the left by $\sigma^{\mu},$ while left-handed fermions are acted upon from the left by $\bar{\sigma}^{\mu}.$
Here, $\bar{\sigma}^0 = \sigma^0$ while $\bar{\sigma}^j = -\sigma^j$ for $j\in\{1,2,3\}.$  

 On the other hand, the hermitian conjugate of a left-handed Weyl spinor is right-handed, and is acted upon from the right with $\bar{\sigma}^{\mu},$ while the hermitian conjugate of a right-handed Weyl spinor is left-handed, and is acted upon from the right with ${\sigma}^{\mu}$, \citep{TASI}.  Thus, the left-handed fermions in our model will have their spin/helicity states identified via the right action of $\sigma^3,$ while the right-handed fermions will have their spin/helicity states identified via the left action of $\sigma^3.$

It can furthermore be seen that transitions between generations in these outer off-diagonal blocks are given by excitations analogous to excitations between spin states.  As will be explained later in Section~\ref{disc}, generations (mass) and spin have a special relationship in this model because they are linked via the SO$^+$(3,1) inner automorphisms of $\CH.$

For the remainder of this article, we will define the basis vectors ${\sigma}^{\mu}$ as
\begin{equation}\begin{array}{lll}
\sigma^0:=v_{\mathbb{H}} + \beta_1^{\dagger}v_{\mathbb{H}}\beta_1 &\hspace{5mm}&
\sigma^1:=\beta_1^{\dagger}v_{\mathbb{H}} + v_{\mathbb{H}}\beta_1 \vspace{2mm}\\
\sigma^2:=i(\beta_1^{\dagger}v_{\mathbb{H}} - v_{\mathbb{H}}\beta_1) &\hspace{5mm}&
\sigma^3:=v_{\mathbb{H}} - \beta_1^{\dagger}v_{\mathbb{H}}\beta_1.
\end{array}\end{equation}
\noindent

We define the total vacuum, $v,$ as $v:=v_{\mathbb{H}}v_{\mathbb{O}} = v_{\mathbb{O}}v_{\mathbb{H}}.$

$$\begin{array}{lll}
\textup{\bf Particle content} &\hspace{.01cm} &\textup{\bf Peirce subspace}  \vspace{2mm}\\
SU(2)_L\textup{-active}  & &\mathcal{H}_{13}  \\
\textup{anti-leptons } (8\hspace{.5mm}\C)&&\vspace{2mm}\\
\left(\mathbf{\underline{1}},\mathbf{\underline{2}},\frac{1}{2}\right) \times 4& & \color{gray}{ \overline{\nu_{eL}}^{\uparrow}}\hspace{.5mm} \color{black}v\alpha_1 + \color{gray}\overline{\nu_{eL}}^{\downarrow} \hspace{.5mm}\color{black}\beta_4^{\dagger}v\alpha_1 \vspace{2mm}\\

& & +\hspace{.5mm}\color{gray}{ \overline{e_L}^{\uparrow}}\hspace{.5mm} \color{black}v\alpha_2 + \color{gray}{ \overline{e_L}^{\downarrow}} \hspace{.5mm}\color{black}\beta_4^{\dagger}v\alpha_2
\vspace{2mm}\\

 & & +\hspace{.5mm}\color{gray}{\overline{\nu_{\mu_L}}^{\uparrow}}\hspace{.5mm} \color{black}v\alpha_1\beta_4 + \color{gray}{ \overline{\nu_{\mu L}}^{\downarrow}} \hspace{.5mm}\color{black}\beta_4^{\dagger}v\alpha_1\beta_4 \vspace{2mm}\\
 
& & +\hspace{.5mm}\color{gray}{\overline{\mu_L}^{\uparrow}}\hspace{.5mm} \color{black}v\alpha_2\beta_4 + \color{gray}{ \overline{{\mu_L}}^{\downarrow}} \hspace{.5mm}\color{black}\beta_4^{\dagger}v\alpha_2\beta_4 \vspace{2mm}\\
&&+\hspace{.5mm} h.c. 

\vspace{5mm}\\

SU(2)_L\textup{-active} & \hspace{.01cm} &\mathcal{H}_{23} \\
 \textup{quarks } (24\hspace{.5mm}\C)\vspace{2mm}\\
 
\left(\mathbf{\underline{3}},\mathbf{\underline{2}},\frac{1}{6}\right) \times 4& & \color{gray}{d_L^{k\downarrow}}\hspace{.5mm} \color{black}\varepsilon_{ijk}\alpha_i^{\dagger}\alpha_j^{\dagger}v\alpha_1 \vspace{2mm}\\

&&+\hspace{.5mm} \color{gray}s_L^{k\downarrow} \hspace{.5mm}\color{black}\varepsilon_{ijk}\beta_4^{\dagger}\alpha_i^{\dagger}\alpha_j^{\dagger}v\alpha_1 \vspace{2mm}\\

& & +\hspace{.5mm}\color{gray}{u_{L}^{k\downarrow}}\hspace{.5mm} \color{black}\varepsilon_{ijk}\alpha_i^{\dagger}\alpha_j^{\dagger}v\alpha_2 \vspace{2mm}\\
&&+\hspace{.5mm} \color{gray}{c_{L}^{k\downarrow}} \hspace{.5mm}\color{black}\varepsilon_{ijk}\beta_4^{\dagger}\alpha_i^{\dagger}\alpha_j^{\dagger}v\alpha_2
\vspace{2mm}\\

 & & +\hspace{.5mm}\color{gray}{d_L^{k\uparrow}}\hspace{.5mm} \color{black}\varepsilon_{ijk}\alpha_i^{\dagger}\alpha_j^{\dagger}v\alpha_1\beta_4 \vspace{2mm}\\
&& +\hspace{.5mm} \color{gray}{s_L^{k\uparrow}} \hspace{.5mm}\color{black}\varepsilon_{ijk}\beta_4^{\dagger}\alpha_i^{\dagger}\alpha_j^{\dagger}v\alpha_1\beta_4 \vspace{2mm}\\
& & +\hspace{.5mm}\color{gray}{u_{L}^{k\uparrow}}\hspace{.5mm} \color{black}\varepsilon_{ijk}\alpha_i^{\dagger}\alpha_j^{\dagger}v\alpha_2\beta_4 \vspace{2mm}\\
&&+\hspace{.5mm} \color{gray}{c_{L}^{k\uparrow}} \hspace{.5mm}\color{black}\varepsilon_{ijk}\beta_4^{\dagger}\alpha_i^{\dagger}\alpha_j^{\dagger}v\alpha_2\beta_4 + h.c.

\vspace{5mm}\\

SU(2)_L\textup{-inactive}  &\hspace{.01cm} &\mathcal{H}_{14}  \\
\textup{anti-neutrinos }\\(4\hspace{.5mm}\C)&&
\vspace{2mm}\\

\left(\mathbf{\underline{1}},\mathbf{\underline{1}},0\right) \times 4& & \color{gray}{ \overline{\nu_{e R}}^{\downarrow}}\hspace{.5mm} \color{black}v\alpha_3 + \color{gray} \overline{\nu_{\mu R}}^{\downarrow} \hspace{.5mm}\color{black}\beta_4^{\dagger}v\alpha_3 \vspace{2mm}\\

 & & +\hspace{.5mm}\color{gray} \overline{\nu_{e R}}^{\uparrow}\hspace{.5mm} \color{black}v\alpha_3\beta_4  + \color{gray}\overline{\nu_{\mu R}}^{\uparrow} \hspace{.5mm}\color{black}\beta_4^{\dagger}v\alpha_3\beta_4\vspace{2mm}\\
 && +\hspace{.5mm} h.c.

\end{array}$$

$$\begin{array}{lll}

SU(2)_L\textup{-inactive}  & \hspace{.01cm} &\mathcal{H}_{24} \\
\textup{down-type quarks } \\
(12\hspace{.5mm}\C) \vspace{2mm}&&\\

\left(\mathbf{\underline{3}},\mathbf{\underline{1}},-\frac{1}{3}\right) \times 4& & \color{gray}{ d_R^{k{\uparrow}}}\hspace{.5mm} \color{black}\varepsilon_{ijk}\alpha_i^{\dagger}\alpha_j^{\dagger}v\alpha_3 \vspace{2mm}\\
&&+\hspace{.5mm} \color{gray} d_R^{k{\downarrow}} \hspace{.5mm}\color{black}\varepsilon_{ijk}\beta_4^{\dagger}\alpha_i^{\dagger}\alpha_j^{\dagger}v\alpha_3 \vspace{2mm}\\

 & & +\hspace{.5mm}\color{gray}{ s_R^{k{\uparrow}}}\hspace{.5mm} \color{black}\varepsilon_{ijk}\alpha_i^{\dagger}\alpha_j^{\dagger}v\alpha_3\beta_4 \vspace{2mm}\\
&& +\hspace{.5mm} \color{gray}{ s_R^{k{\downarrow}}} \hspace{.5mm}\color{black}\varepsilon_{ijk}\beta_4^{\dagger}\alpha_i^{\dagger}\alpha_j^{\dagger}v\alpha_3\beta_4 \vspace{2mm}\\
&&+\hspace{.5mm} h.c.

    \vspace{5mm}\\

 SU(2)_L\textup{-inactive}  &\hspace{.01cm} &\mathcal{H}_{15}  \\
\textup{charged anti-leptons }(4\hspace{.5mm}\C)
\vspace{2mm}\\

\left(\mathbf{\underline{1}},\mathbf{\underline{1}},1\right) \times 4 & & \color{gray}  \overline{{e_R}}^{\downarrow}\hspace{.5mm} \color{black}v\alpha_1\alpha_2\alpha_3 \vspace{2mm}\\
&&+\hspace{.5mm} \color{gray} \overline{\mu_R}^{\downarrow} \hspace{.5mm}\color{black}\beta_4^{\dagger}v\alpha_1\alpha_2\alpha_3
\vspace{2mm}\\

& & +\hspace{.5mm}\color{gray} \overline{e_R}^{\uparrow}\hspace{.5mm} \color{black}v\alpha_1\alpha_2\alpha_3\beta_4 \vspace{2mm}\\
&&+\hspace{.5mm} \color{gray} \overline{\mu_R}^{\uparrow} \hspace{.5mm}\color{black}\beta_4^{\dagger}v\alpha_1\alpha_2\alpha_3\beta_4 + h.c.

  \vspace{5mm}\\

SU(2)_L\textup{-inactive}  &\hspace{.01cm} &\mathcal{H}_{25}  \\
\textup{up-type quarks }\\
(12\hspace{.5mm}\C) &&
\vspace{2mm}\\
\left(\mathbf{\underline{3}},\mathbf{\underline{1}},\frac{2}{3}\right)\times 4 & & \color{gray}{ u_R^{k\uparrow}}\hspace{.5mm} \color{black}\varepsilon_{ijk}\alpha_i^{\dagger}\alpha_j^{\dagger}v\alpha_1\alpha_2\alpha_3 \vspace{2mm}\\
&&+\hspace{.5mm} \color{gray}{  u_R^{k{\downarrow}}} \hspace{.5mm}\color{black}\varepsilon_{ijk}\beta_4^{\dagger}\alpha_i^{\dagger}\alpha_j^{\dagger}v\alpha_1\alpha_2\alpha_3
\vspace{2mm}\\

& & +\hspace{.5mm}\color{gray}{ c_R^{k{\uparrow}}}\hspace{.5mm} \color{black}\varepsilon_{ijk}\alpha_i^{\dagger}\alpha_j^{\dagger}v\alpha_1\alpha_2\alpha_3\beta_4 \vspace{2mm}\\
&&+\hspace{.5mm} \color{gray}{c_{R}^{k\downarrow}} \hspace{.5mm}\color{black}\varepsilon_{ijk}\beta_4^{\dagger}\alpha_i^{\dagger}\alpha_j^{\dagger}v\alpha_1\alpha_2\alpha_3\beta_4 \vspace{2mm}\\
&&+\hspace{1mm} h.c.
\end{array}$$
It should be noted that we have labeled fermions in this section indicating a specific choice of generation.  This is merely for notational convenience; it is best not to commit yet to family identification.

\newpage

\subsubsection{Diagonal blocks  \\Covariant derivative \\$64\hspace{.5mm}\mathbb{R}$}

\begin{figure}[h]
\begin{center}
\includegraphics[width=7cm]{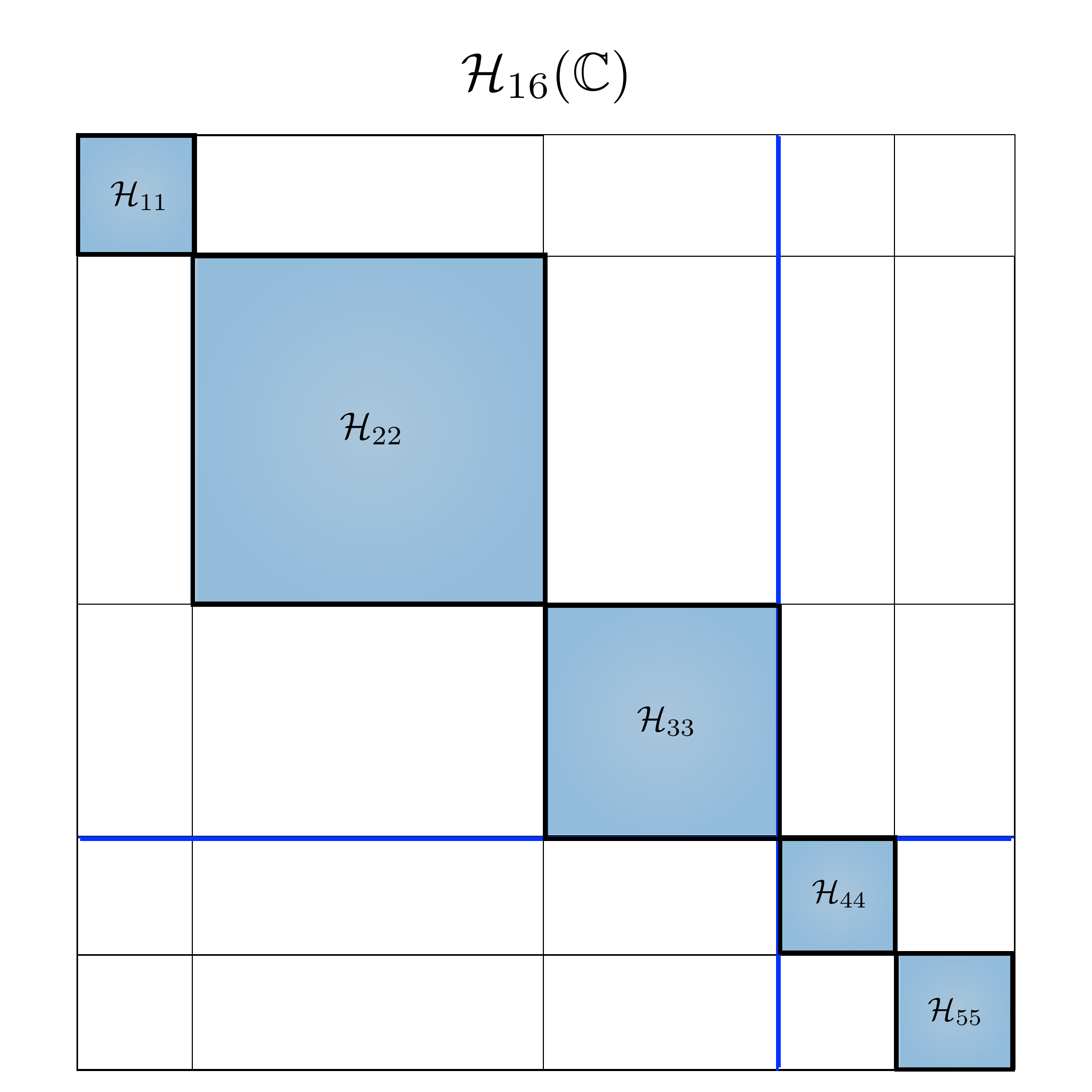}
\caption{\label{diagfig}  Momentum-space covariant derivative in the diagonal blocks }
\end{center}\end{figure}

Given that diagonal blocks transform under the adjoint representation, it is here that a momentum space covariant derivative may be identified.  Please see Figure~(\ref{diagfig}).  Interestingly, this covariant derivative fragments according to the Peirce decomposition as 
acting on (colourless or colourful particles) and (SU(2)$_L$-active or -inactive particles).

$$\begin{array}{lll}
\textup{\bf Particle content} &\hspace{.2cm} &\textup{\bf Peirce subspace}  \vspace{2mm}\\
\textup{Colourless covariant derivative} & &\mathcal{H}_{11}  \vspace{2mm}\\
p_{\mu}^{\ell}\sigma^{\ell\mu} \hspace{2mm} (4\hspace{.5mm}\mathbb{R}) & &\color{gray}{p_{\mu}^{\ell}}\hspace{.5mm} \color{black} s_1 \sigma^{\mu} \vspace{2mm}\\

\left(\mathbf{\underline{1}},\mathbf{\underline{1}},0\right) \times 4& & 

 \vspace{5mm}\\

\textup{Colourful covariant derivative} &\hspace{.2cm} &\mathcal{H}_{22} \vspace{2mm}\\
p_{\mu}^{q}\sigma^{q\mu}  \hspace{2mm} (4\hspace{.5mm}\mathbb{R}) & &\color{gray}{p_{\mu}^{q}}\hspace{.5mm}  \color{black} s_2\sigma^{\mu}\vspace{2mm}\\ 
\left(\mathbf{\underline{1}},\mathbf{\underline{1}},0\right) \times 4& & \vspace{2mm}\\

\textup{Gluons } (32\hspace{.5mm}\mathbb{R}) & &  \color{gray}{G^j_{\mu}}\hspace{.5mm} \color{black} s_2 \Lambda_j\sigma^{\mu}\vspace{2mm}\\
\left(\mathbf{\underline{8}},\mathbf{\underline{1}},0\right) \times 4& &



  \vspace{5mm}\\

\textup{SU(2)$_L$-active covariant derivative} & \hspace{.2cm}&\mathcal{H}_{33}\vspace{2mm}\\
p_{\mu}^{a}\sigma^{a\mu} \hspace{2mm} (4\hspace{.5mm}\mathbb{R}) & & \color{gray}{p_{\mu}^{a}}\hspace{.5mm} \color{black} s_3 \sigma^{\mu} \vspace{2mm}\\
\left(\mathbf{\underline{1}},\mathbf{\underline{1}},0\right) \times 4& &  \vspace{2mm}\\
W_L \textup{ bosons } (12\hspace{.5mm}\mathbb{R}) & & \color{gray}{W^m_{L\mu}}\hspace{.5mm} \color{black}  s_3\tau_m\sigma^{\mu} \vspace{2mm}\\
\left(\mathbf{\underline{1}},\mathbf{\underline{3}},0\right) \times 4& &

  
   \end{array}$$

 $$\begin{array}{lll}

\textup{SU(2)$_L$-inactive covariant derivative} &\hspace{.2cm} &\mathcal{H}_{44}+\mathcal{H}_{55}\vspace{2mm} \\
p_{\mu}^{\bar{a}}\sigma^{\bar{a}\mu}  \hspace{2mm} (4\hspace{.5mm}\mathbb{R}) & &\color{gray}{p_{\mu}^{\bar{a}}}\hspace{.5mm} \color{black} \left(s_4 +s_5\right)\sigma^{\mu} \vspace{2mm}\\
\left(\mathbf{\underline{1}},\mathbf{\underline{1}},0\right) \times 4& & \vspace{2mm}\\
W^3_R \textup{ component } (4\hspace{.5mm}\mathbb{R}) & &  \color{gray}{W^3_{R\mu}}\hspace{.5mm} \color{black}  \left(s_4 -s_5\right)\sigma^{\mu} \vspace{2mm}\\
\left(\mathbf{\underline{1}},\mathbf{\underline{1}},0\right) \times 4& &
\end{array}$$

As with the fermion representations of the previous section, the gauge bosons represented here are seen to be off-shell (unconstrained).

Unlike with the gluons and W bosons represented above, the gauge boson associated with weak hypercharge, $B_{\mu},$ spans across several Peirce blocks.  In fact, it is the only Standard Model particle representation that does so.  Explicitly, it is given by linear combinations of the above degrees of freedom,
\begin{equation}
\color{gray}B_{\mu}\color{black} \hspace{.5mm}\frac{1}{2}(s_1+\frac{1}{3}s_2+s_4-s_5) \hspace{.5mm}\sigma^{\mu}.
\end{equation}
\noindent The fact that it spans across several Peirce blocks, as opposed to residing within a single Peirce block might be of concern, at first sight.  That is, our original intension was to have particle representations fit within the Peirce blocks.  However, as may be confirmed in Section~\ref{Z2}, one finds that the set of all diagonal blocks together form a subspace of the \it same grade. \rm That is, the $B_{\mu}$ boson, and all particles in this model for that matter, indeed respects the $\mathbb{Z}_2^5$ grading.

Readers should notice that this model has not yet been gauged.  Of course, gauge bosons are known to transform infinitesimally as $\delta G^a_{\mu} = gF^{abc}\theta^bG^c_{\mu} + \partial _{\mu}\theta^a,$ where $g$ is a coupling constant, $F^{abc}$ are the Lie algebraic structure constants, and $\theta^a$ are  local parameters.  In other words, a gauge boson's infinitesimal transformation comprises  an adjoint term, and an inhomogeneous term, respectively.  At the moment, the diagonal blocks of $\mathcal{H}_{16}(\C)$ presented here transform only under the adjoint representation.  We will likely need to consider \it sequences \rm of $\mathcal{H}_{16}(\C)$ elements under multiplication in order to construct the canonical transformation of gauge fields, including finally  their inhomogeneous term.   It is for this reason that we refer to the diagonal blocks in this model as a ``cursory" covariant derivative.

\vspace{5cm}

\subsubsection{Inner off-diagonal blocks   \\Third generation  \\ $64\hspace{.5mm}\mathbb{R}$\label{inoff}}

\begin{figure}[h!!]
\begin{center}
\includegraphics[width=7cm]{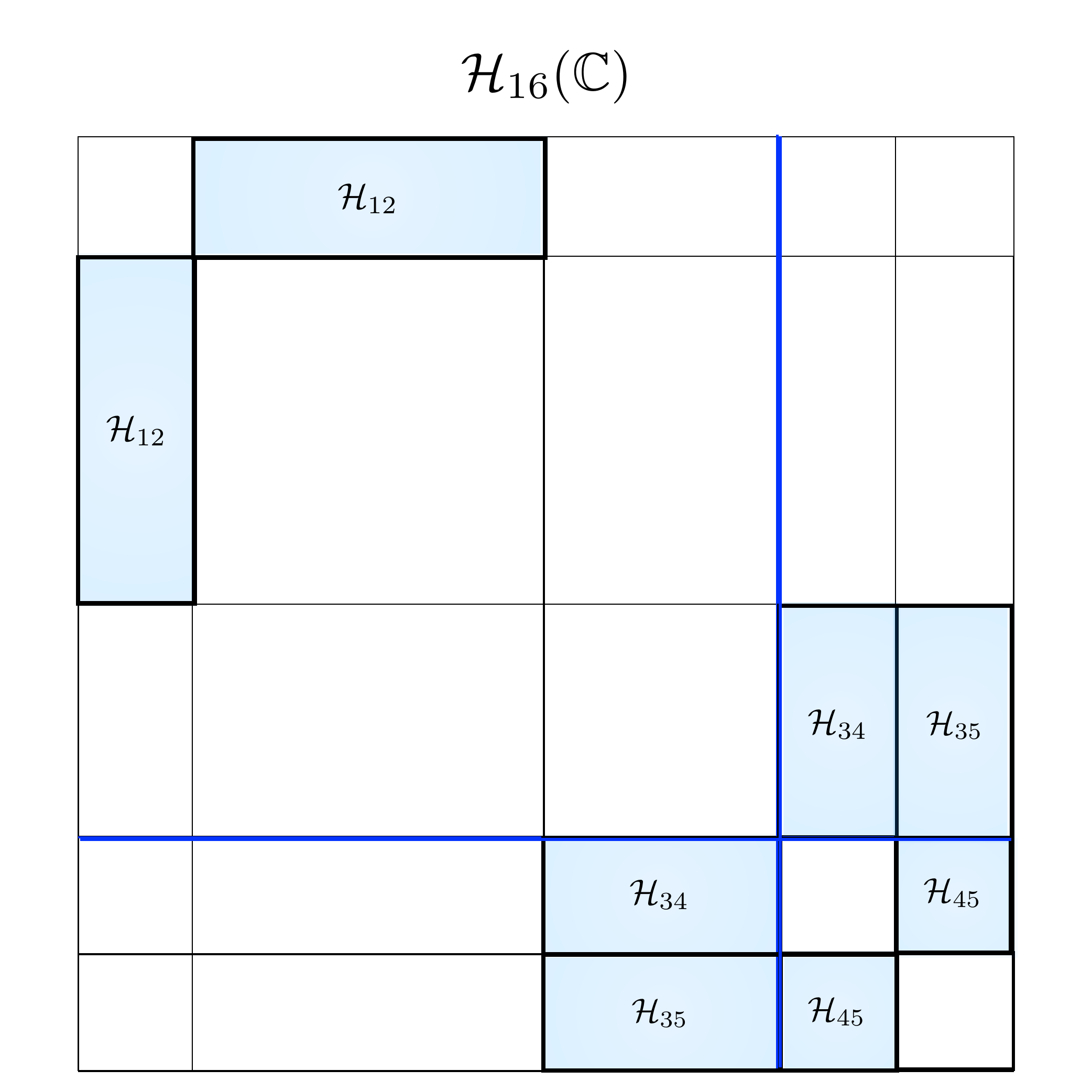}
\caption{\label{inneroff}  Third generation in the inner off-diagonal blocks }
\end{center}\end{figure}

Least trivial are the final $64\hspace{.5mm}\mathbb{R}$ degrees of freedom, given by the inner off-diagonal blocks.  Please see Figure~(\ref{inneroff}).  Had our Peirce decomposition been given by a coarse grained set $\mathcal{S}_2:=\{s, s^*\},$ then we would interpret these representations as bosonic.  In this case, our model would indeed have a 1:1 boson-to-fermion ratio.  

However, from the $\mathcal{S}_5$ perspective, one finds a set of representations, which, in every case, \it aligns with a third generation fermion. \rm  In fact, one finds particle irreps for each of the lightest third generation states.  This sector is seen to be missing two known Standard Model irreps, namely, $(t_L, b_L)$ and $t_R$.  These  coincide exactly with those irreps needed to describe the top quark.

For the expressions below, let $i,j,k \in\{1,2,3\}$.

$$\begin{array}{lll}
\textup{\bf Particle content} &\hspace{.5cm} &\textup{\bf Peirce subspace}  \vspace{2mm}\\
\textup{SU(2)$_L$-inactive} & &\mathcal{H}_{12}  \\
\textup{anti-bottom quarks } (12\hspace{.5mm}\C)\vspace{2mm}\\
\left(\mathbf{\underline{3}^*},\mathbf{\underline{1}},\frac{1}{3}\right) \times 4& & \color{gray}{\overline{b_R}^{1k\downarrow}} \color{black}\varepsilon_{ijk}\hspace{.5mm}v_{\mathbb{O}}\hspace{.5mm}\alpha_i\alpha_j \vspace{2mm}\\
&& \color{gray}{\overline{b_R}^{2k\downarrow}} \color{black}\varepsilon_{ijk}\hspace{.5mm}\beta_4^{\dagger}v_{\mathbb{O}}\hspace{.5mm}\alpha_i\alpha_j
 \vspace{2mm}\\
&& \color{gray}{\overline{b_R}^{1k\uparrow}} \color{black}\varepsilon_{ijk}\hspace{.5mm}v_{\mathbb{O}}\hspace{.5mm}\alpha_i\alpha_j\beta_4
\vspace{2mm}\\
&& \color{gray}{\overline{b_R}^{2k\uparrow}} \color{black}\varepsilon_{ijk}\hspace{.5mm}\beta_4^{\dagger}v_{\mathbb{O}}\hspace{.5mm}\alpha_i\alpha_j\beta_4
\vspace{2mm}\\

&&+\hspace{.5mm}h.c.

 \end{array}$$

$$\begin{array}{lll}

\textup{SU(2)$_L$-active anti-leptons} & \hspace{.15cm} &\mathcal{H}_{35}  \\
  (8\hspace{.5mm}\C)\vspace{2mm}\\
\left(\mathbf{\underline{1}},\mathbf{\underline{2}},\frac{1}{2}\right) \times 4& & \color{gray}{\overline{\tau_L}^{1\uparrow}}\hspace{.5mm} \color{black}\alpha^{\dagger}_1v\alpha_1\alpha_2\alpha_3  \vspace{2mm} \\
& & +\hspace{.5mm}\color{gray}{\overline{\tau_L}^{1\downarrow}}\hspace{.5mm} \color{black}\beta_4^{\dagger}\alpha^{\dagger}_1v\alpha_1\alpha_2\alpha_3  \vspace{2mm} \\
& & +\hspace{.5mm}\color{gray}{\overline{\tau_L}^{2\uparrow}}\hspace{.5mm} \color{black}\alpha^{\dagger}_1v\alpha_1\alpha_2\alpha_3\beta_4  \vspace{2mm} \\
& & +\hspace{.5mm}\color{gray}{\overline{\tau_L}^{2\downarrow}}\hspace{.5mm} \color{black}\beta_4^{\dagger}\alpha^{\dagger}_1v\alpha_1\alpha_2\alpha_3\beta_4   \vspace{2mm} \\

& &+\hspace{.5mm} \color{gray}{\overline{\nu_{\tau L}}^{1\uparrow}}\hspace{.5mm} \color{black}\alpha^{\dagger}_2v\alpha_1\alpha_2\alpha_3  \vspace{2mm} \\
& & +\hspace{.5mm}\color{gray}{\overline{\nu_{\tau L}}^{1\downarrow}}\hspace{.5mm} \color{black}\beta_4^{\dagger}\alpha^{\dagger}_2v\alpha_1\alpha_2\alpha_3  \vspace{2mm} \\
& & +\hspace{.5mm}\color{gray}{\overline{\nu_{\tau L}}^{2\uparrow}}\hspace{.5mm} \color{black}\alpha^{\dagger}_2v\alpha_1\alpha_2\alpha_3\beta_4  \vspace{2mm} \\
& & +\hspace{.5mm}\color{gray}{\overline{\nu_{\tau L}}^{2\downarrow}}\hspace{.5mm} \color{black}\beta_4^{\dagger}\alpha^{\dagger}_2v\alpha_1\alpha_2\alpha_3\beta_4  
\vspace{2mm} \\

&&+\hspace{.5mm} h.c.

 \vspace{5mm}\\
 


\textup{SU(2)$_L$-inactive anti-tau} & \hspace{.15cm} &\mathcal{H}_{45} \\
  (4\hspace{.5mm}\C)\vspace{2mm}\\
\left(\mathbf{\underline{1}},\mathbf{\underline{1}},1\right) \times 4& & \color{gray}{\overline{\tau_R}^{1\downarrow}}\hspace{.5mm} \color{black}\alpha^{\dagger}_3v\alpha_1\alpha_2\alpha_3 \vspace{2mm} \\
& & +\hspace{.5mm}\color{gray}{\overline{\tau_R}^{2\downarrow}}\hspace{.5mm} \color{black}\beta_4^{\dagger}\alpha^{\dagger}_3v\alpha_1\alpha_2\alpha_3  \vspace{2mm} \\
& & +\hspace{.5mm}\color{gray}{\overline{\tau_R}^{1\uparrow}}\hspace{.5mm} \color{black}\alpha^{\dagger}_3v\alpha_1\alpha_2\alpha_3\beta_4  \vspace{2mm} \\
& & +\hspace{.5mm}\color{gray}{\overline{\tau_R}^{2\uparrow}}\hspace{.5mm} \color{black}\beta_4^{\dagger}\alpha^{\dagger}_3v\alpha_1\alpha_2\alpha_3\beta_4 \vspace{2mm} \\

&&+\hspace{.5mm} h.c.

\end{array}$$

At this point, we have identified all third generation representations, with exception to those irreps involving the top quark, and also a possible third generation sterile neutrino.  We are furthermore missing the Higgs.  On the other hand, the final Peirce block we have yet to describe is $\mathcal{H}_{34}$.  This $\mathcal{H}_{34}$ is unique within the Peirce decomposition because it is the only block that is bosonic with respect to $\mathcal{S}_4,$ and fermionic with respect to $\mathcal{S}_5.$

$$\begin{array}{lll}

\textup{Higgs} \cdot \textup{(anti)-neutrino} & \hspace{.5cm} &\mathcal{H}_{34}  \\
 (8\hspace{.5mm}\C)\vspace{2mm}\\
\left(\mathbf{\underline{1}},\mathbf{\underline{2}},-\frac{1}{2}\right) \times 4& & \color{gray}{ h^{\uparrow 1}\overline{\nu_{\tau R}}^{\downarrow}}\hspace{.5mm} \color{black}\alpha^{\dagger}_1v\alpha_3  \vspace{2mm} \\
& & +\hspace{.5mm}\color{gray}{ h^{\uparrow 2}\overline{\nu_{\tau R}}^{\downarrow}}\hspace{.5mm} \color{black}\beta_4^{\dagger}\alpha^{\dagger}_1v\alpha_3  \vspace{2mm} \\
& & +\hspace{.5mm}\color{gray}{ h^{\uparrow 1}\overline{\nu_{\tau R}}^{\uparrow}}\hspace{.5mm}\color{black}\alpha^{\dagger}_1v\alpha_3\beta_4  \vspace{2mm} \\

& & +\hspace{.5mm}\color{gray}{ h^{\uparrow 2}\overline{\nu_{\tau R}}^{\uparrow}}\hspace{.5mm} \color{black}\beta_4^{\dagger}\alpha^{\dagger}_1v\alpha_3\beta_4  \vspace{2mm} \\

 & & +\hspace{.5mm}\color{gray}{ h^{\downarrow 1}\overline{\nu_{\tau R}}^{\downarrow}}\hspace{.5mm} \color{black}\alpha^{\dagger}_2v\alpha_3  \vspace{2mm} \\
 
& & +\hspace{.5mm}\color{gray}{ h^{\downarrow 2}\overline{\nu_{\tau R}}^{\downarrow}}\hspace{.5mm} \color{black}\beta_4^{\dagger}\alpha^{\dagger}_2v\alpha_3  \vspace{2mm} \\

& & +\hspace{.5mm}\color{gray}{ h^{\downarrow 1}\overline{\nu_{\tau R}}^{\uparrow}}\hspace{.5mm}\color{black}\alpha^{\dagger}_2v\alpha_3\beta_4  \vspace{2mm} \\
& & +\hspace{.5mm}\color{gray}{ h^{\downarrow 2}\overline{\nu_{\tau R}}^{\uparrow}}\hspace{.5mm} \color{black}\beta_4^{\dagger}\alpha^{\dagger}_2v\alpha_3\beta_4  \vspace{2mm} \\

&&+\hspace{.5mm} h.c.
\end{array}$$

We observe that this final Peirce block transforms as a product of the Higgs and sterile anti-neutrino (or neutrino), and so we will take the liberty to describe it here in that way.  It should be noted that $\mathcal{H}_{34}$ and $\mathcal{H}_{35}$ correspond to essentially the same representation, and so there may be some freedom of interpretation for these two Peirce blocks.

This now completes the particle identification of our Peirce blocks.  One finds, curiously, that the light third generation fermions are described in duplicate.  It is not clear whether or not this trait need be detectable.   

So, how might one account for the existence of these heaviest top and bottom quarks?  On one hand, it is plausible that the top quark could be composite, while the bottom quark could be partially composite.  Indeed, there is precedence for such proposals,~\citep{kaplan},  \citep{tqc}, \citep{topcomp}.  

As an alternative perspective, one might postulate that the third generation states could ultimately enter as a product, reminiscent of division algebraic triality, proposed in~\citep{TTT}.  In~\citep{TTT}, the Standard Model's three generations were described as a triality triple, with the vector representation entering in as products of third generation spinors $\sim \sum \Psi_{\textup{III}} \Psi_{\textup{III}}'^{\dagger}.$  Following this example, we would then relabel the inner off-diagonal states in the following way:

\begin{equation}\begin{array}{l}\label{factor}
\textup{Anti-bottom quark }\hspace{.5mm} \left(\mathbf{\underline{3}^*},\mathbf{\underline{1}},\frac{1}{3}\right) \times 4  \hspace{.5mm}+ \hspace{.5mm} h.c. \vspace{2mm}\\

\hspace{1cm}\mapsto \hspace{.5cm}\overline{\mathcal{L}_L}^{\textup{III}} \hspace{.5mm}\mathcal{Q}_L^{\textup{III}\dagger} +  \overline{\tau_R}\hspace{.5mm}t_R^{\dagger}+ \overline{\nu_{\tau_R}}\hspace{.5mm}b_R^{\dagger}\hspace{.5mm}+ \hspace{.5mm} h.c.  \vspace{5mm}\\

\textup{SU(2)$_L$-active anti-leptons }\hspace{.5mm} \left(\mathbf{\underline{1}},\mathbf{\underline{2}},\frac{1}{2}\right) \times 4 \hspace{.5mm}+ \hspace{.5mm} h.c. \vspace{2mm}\\
\hspace{1cm}\mapsto \hspace{.5cm}\overline{\mathcal{L}_L}^{\textup{III}\dagger} \hspace{.5mm}\overline{\tau_R} + \mathcal{Q}_L^{\textup{III}\dagger} \hspace{.5mm}t_R  \hspace{.5mm}+ \hspace{.5mm} h.c. \vspace{5mm}\\

\textup{SU(2)$_L$-inactive anti-tau }\hspace{.5mm} \left(\mathbf{\underline{1}},\mathbf{\underline{1}},1\right) \times 4 \hspace{.5mm}+ \hspace{.5mm} h.c. \vspace{2mm}\\
\hspace{1cm}\mapsto \hspace{.5cm}\overline{\nu_{\tau R}}^{\dagger}  \hspace{.5mm}\overline{\tau_R}+ b_R^{\dagger} t_R  \hspace{.5mm}+ \hspace{.5mm} h.c. 
\vspace{5mm}\\

\textup{Higgs} \cdot \textup{(anti)-neutrino} \hspace{.5mm}\left(\mathbf{\underline{1}},\mathbf{\underline{2}},-\frac{1}{2}\right) \times 4 \hspace{.5mm}+ \hspace{.5mm} h.c. \vspace{2mm}\\
\hspace{1cm}\mapsto \hspace{.5cm}\overline{\mathcal{L}_L}^{\textup{III}\dagger} \hspace{.5mm}\overline{\nu_{\tau_R}} + \mathcal{Q}_L^{\textup{III}\dagger} \hspace{.5mm}b_R  \hspace{.5mm}+ \hspace{.5mm} h.c.

\end{array}\end{equation}

\noindent This provides one way for the full set of third generation states, including top and bottom quarks, to embed into the algebra, albeit as a product.  To a certain extent, it mirrors structure native to the spinor-helicity formalism.  

These particular combinations of products were chosen specifically because they align with the way that the first and second generation fermions mix in under Jordan algebraic multiplication.  Worth noting here is the fact that it is possible in this model to multiply two fermions in order to obtain another fermion.  This is a direct consequence of our definition of fermions as off-diagonal elements within the Peirce decomposition.  For more on this topic, please see Section~\ref{Z2}.

\subsubsection{A Jordan algebraic mosaic \label{full}}

\begin{figure*}[ht]
\begin{center}
\includegraphics[width=.65\textwidth]{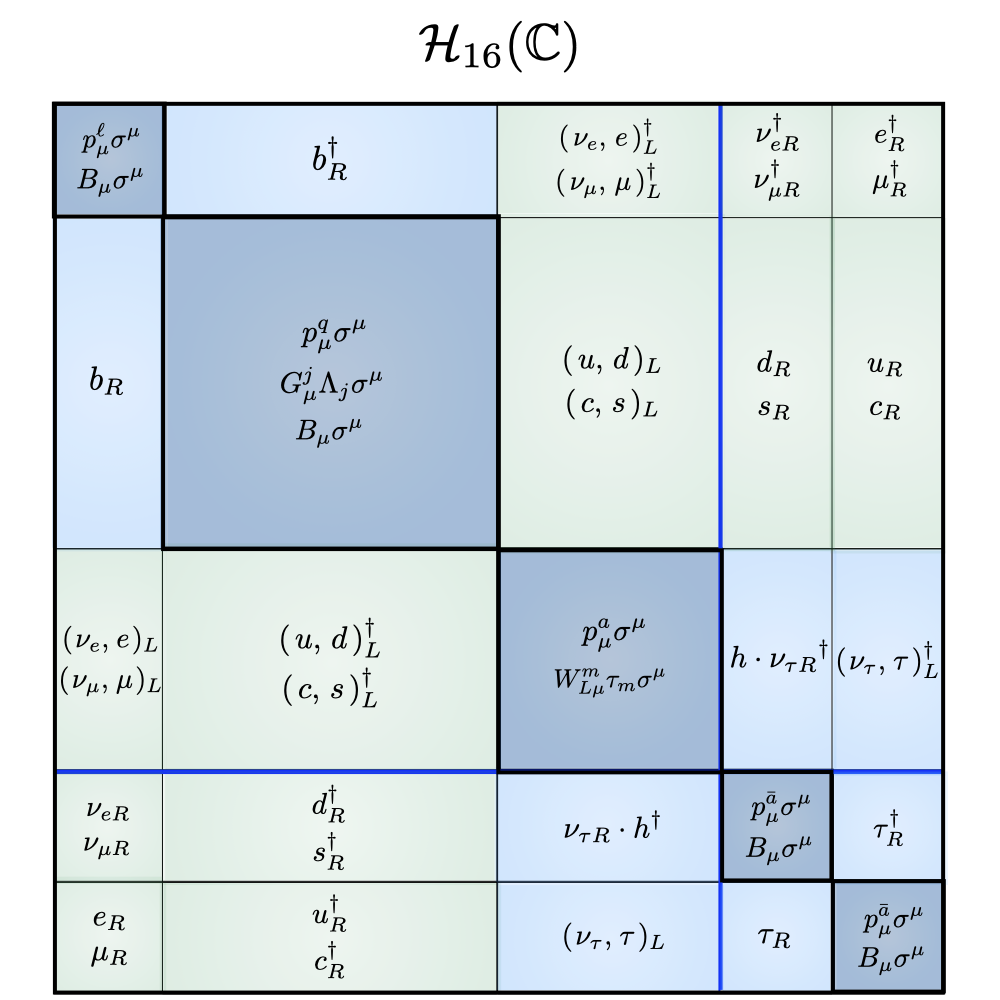}
\caption{\label{map}  Peirce decomposition of $\mathcal{H}_{16}(\C)$ into familiar covariant derivative and fermionic vector spaces.     \\Note:  (1) Fermions are identified as off-diagonal objects within the Peirce decomposition.  Physically, we say that they extend between two distinct sites, $s_i$ and $s_j$, each subject to an independent $\mathfrak{su}(2)$ transformation.  This is a fundamentally different description than that of the Standard Model.  Please see Section~\ref{disc} for details.  (2) Representations $(\nu_{\tau}, \tau)_L$ and $h\cdot \nu_{\tau R}^{\dagger}$ are identical under $\gsm$.  For visual simplicity we identify each with a particular Peirce block.  Also mostly for visual simplicity, we have labeled Peirce blocks with explicit generation labels.  (3)  A certain sign change within the weak hypercharge operator allows for the interchange of RH neutrino and charged lepton so that neutrinos may be positioned in the extreme $\mathcal{H}_{15}$ corners of the diagram.  This was depicted in the first arXiv version of the article.}
\end{center}\end{figure*}

In summary, within $\mathcal{H}_{16}(\C)$ we find irreducible representations corresponding to a (cursory) momentum-space covariant derivative, including gluons and electroweak bosons on the diagonal.  We find two full generations of fermions within the outer off-diagonals blocks.  Finally, within the inner-off-diagonal blocks, we find the fermion representations for a third generation, with only the heaviest particles evading a simple description. We identify a possible inclusion of a third sterile neutrino and the Higgs.   Please see Figure~\ref{map}.  

It is worth noting a couple of interesting ratios that arise in counting the Standard Model's states.  In this model, we have found a 3:1 ratio of the fermioinic sector relative to the boson sector.  This counting was made possible in part by the inclusion of separate Weyl operators $p_{\mu}\sigma^{\mu}$ for the various particle representations.  Recently, it was pointed out by Jorge Zanelli that when the Higgs and Weyl operators are excluded, the fermionic ratio becomes 4:1.  

In contrast to the Standard Model's usual counting, we find that the Weyl operators $p_{\mu}\sigma^{\mu}$ appear as distinct degrees of freedom.  That is before any constraints are applied to the $\mathcal{H}_{16}(\C)$ algebra.  It is worth investigating whether these additional Weyl operators may in fact be \it dark matter candidates. \rm

\section{A $\mathbb{Z}_2^5$-graded algebra\label{Z2}}

At this point, we see how $\mathcal{H}_{16}(\C)$ provides a representation space for $\gsm$ and $\mathfrak{so}(3)$ symmetries. \it But in what sense is this a superalgebra? \rm  

Thanks to the Peirce decomposition earlier described, the algebra $\mathcal{H}_{L_{\A}},$ together with $\mathcal{S}_5,$ facilitate not only a $\mathbb{Z}_2$-graded algebra, but furthermore, a $\mathbb{Z}_2^5$-graded algebra.  

We will define a $\mathbb{Z}_2^n$-graded algebra as an algebra $A$ with vector space decomposition 
\begin{equation}
A = \underset{x\in\mathbb{Z}_2^n}\bigoplus A_{x}
\end{equation}
\noindent such that for all $a_x\in A_x$ and $a_y\in A_y,$ multiplication $m$ obeys
\begin{equation}
m(a_x, a_y) \in A_{x+y}.
\end{equation}
\noindent Addition  $x+y$ is performed componentwise, modulo 2.

In particular, a ($\mathbb{Z}_2$-graded) superalgebra can be achieved, as originally claimed, when we coarse grain $\mathcal{S}_5$ to a simpler set defined by $\mathcal{S}_2:=\{\hspace{.5mm}s, \hspace{.5mm}s^*\hspace{.5mm}\},$ for example.  According to this grading, one may identify $A_0$ as bosonic with 
\begin{equation}\begin{array}{lll}
A_0&=&\mathcal{H}_{11}\oplus \mathcal{H}_{12}\oplus \mathcal{H}_{22}\oplus \mathcal{H}_{33}\vspace{2mm}\\
&&\hspace{.5mm}\oplus \hspace{.5mm}\mathcal{H}_{34}\oplus \mathcal{H}_{35}\oplus \mathcal{H}_{44}\oplus \mathcal{H}_{45}\oplus \mathcal{H}_{55}, 
\end{array}\end{equation}
\noindent and $A_1$ as fermionic with 
\begin{equation}\begin{array}{lll}
A_1&=&\mathcal{H}_{13}\oplus \mathcal{H}_{14}\oplus \mathcal{H}_{15}\vspace{2mm}\\
&&\hspace{.5mm} \oplus\hspace{.5mm} \mathcal{H}_{23}  \oplus \mathcal{H}_{24}\oplus \mathcal{H}_{25}. 
\end{array}\end{equation}

In contrast to $\mathcal{S}_2,$ the $\mathcal{S}_5$ case finally enables a 3:1 fermion to boson ratio.  With this fine-grained Peirce decomposition, the diagonal blocks constitute
\begin{equation} A_{(0,0,0,0,0)}=\mathcal{H}_{11}\oplus \mathcal{H}_{22}\oplus \mathcal{H}_{33}\oplus \mathcal{H}_{44}\oplus \mathcal{H}_{55},
\end{equation}
\noindent whereas each off-diagonal $\mathcal{H}_{ij}$ with $i\neq j$ is identified as
\begin{equation}A_{x_{ij}}=\mathcal{H}_{ij},
\end{equation}
\noindent where $x_{ij}$ has $1$s in its $i$th and $j$th entries, and $0$s elsewhere.

Readers may appreciate an interesting feature of this model.  Namely,  a particle's identification as fermionic or bosonic depends on which Peirce decomposition is enabled.  One might even postulate that in the early universe, the Peirce decomposition was more coarse grained, as in $\mathcal{S}_2,$ and bosonic degrees of freedom had a larger presence.

As a particularly relevant sample calculation, consider a generic element, ${P}$ $\in$ $A_{(0,0,0,0,0)},$ acting on a generic element, $\mathcal{Q}_L$ $\in$ $A_{(0,1,1,0,0)}=\mathcal{H}_{23}.$  Under multiplication 
\begin{equation}m({P}, \mathcal{Q}_L) = {P}\mathcal{Q}_L+ \mathcal{Q}_L{P},
\end{equation}
\noindent one finds the action of the momentum space covariant derivative on two generations of SU(2)$_L$-active quarks.  Namely,
\begin{equation}\begin{array}{l}
m(P, \mathcal{Q}_L) = \vspace{2mm}\\

\hspace{7mm}\left(p_{\mu}^{q}  \sigma^{\mu} s_2+ \frac{g_3}{2}G_{\mu}^{j}\Lambda_j\sigma^{\mu} s_2+ \frac{g_1}{12}B_{\mu}\sigma^{\mu}s_2\right)s_2\mathcal{Q}_Ls_3 \vspace{2mm}\\
\hspace{7mm}+ \hspace{.5mm}s_2\mathcal{Q}_Ls_3 \left(p_{\mu}^{L} \sigma^{\mu} s_3 + \frac{g_2}{2}W_{L\mu}^{m}\tau_m  \sigma^{\mu}s_3\right) + h.c.,
\end{array}\end{equation}
\noindent whose output resides also in $A_{(0,1,1,0,0)}.$  

As a further calculation, readers may verify that gluon and W boson represenations may be built by multiplying, for example, two generic elements $\mathcal{Q}_L,$ $\mathcal{Q}_L'$  $\in A_{(0,1,1,0,0)}.$   Of course, $m(\mathcal{Q}_L, \mathcal{Q}_L')\in A_{(0,0,0,0,0)}.$

As a final calculation, interested readers may verify the mixing of generations under multiplication in $\mathcal{H}_{16}(\C).$

\section{Peirce projective measurements and an algebra of observables \label{PVM}}

Physicists may recognize immediately the similarity between a  set of Peirce idempotents, $\mathcal{S}_n,$ and the familiar projective measurement of quantum theory (a projection-valued measure). A \emph{projective measurement} is also described by a set of orthogonal projection operators that sum to one.  

It is straightforward to see that the set of idempotents $\mathcal{S}_5$ may be fine grained further by specifying one quaternionic imaginary unit (Clifford volume element) for each block.  That is, $\mathcal{S}_5=\{sS, sS^*, \dots \}$ may be fine-grained to become $\mathcal{S}_{10}:=\{sS \frac{1}{2}(1+i\epsilon_j), sS \frac{1}{2}(1-i\epsilon_j)\},sS^* \frac{1}{2}(1+i\epsilon_{j'}), sS^* \frac{1}{2}(1-i\epsilon_{j'}), \dots \}$.  Such a set would allow for a measurement of helicity (spin).  As an aside, we mention that a set of such idempotents generates a poset under addition so that coarse-grained projection operators may be constructed from fine-grained ones.

Beyond projective measurements, one finds that a certain \it algebra of observables, \rm $\mathcal{A},$ follows naturally from  $\mathcal{S}_n.$  That is, consider the vector  space $\mathcal{A}$ given by all real linear combinations of the idempotents $s_k$ in $\mathcal{S}_n$.  
\begin{equation}
a = r_k s_k, \hspace{.5cm} \textup{for } a\in\mathcal{A}  \textup{ and } r_k\in \mathbb{R}.
\end{equation}
Elements in this vector space close under Jordan multiplication.  Hence, we identify $\mathcal{A}$ as a certain local algebra of observables, \citep{FR2022}, identifiable as the \it superselection \rm or \it classical observables \rm of~\citep{Giulini}. 

Now, readers may appreciate that the choice of imaginary units used in $\mathcal{S}_5$ and $\mathcal{S}_{10}$ were not unique.   Although we have arbitrarily chosen $e_7$ to define $s$ and $S,$ octonionic imaginary units reside on the six-sphere, $S^6,$ and so $\mathcal{S}_n$ may be expected to vary along different points in an emergent spacetime.  Similarly, quaternionic imaginary units reside on the two-sphere, $S^2.$  See for example, \citep{mosaic}, \citep{FR2022}, and recent work by Hun Jang,~\citep{Jang}.

This algebra of observables, $\mathcal{A}$,  is of particular importance in this article.  That is, readers should notice that the Peirce decompositions $\mathcal{S}_5$ and $\mathcal{S}_{10},$ derived from division algebraic volume elements,  \it preclude colour as an observable quantity. \rm  

We propose investigating the idea that $\mathcal{A}(x),$ supported by a suitably-defined connection, could comprise a full algebra of observables for the system.

\section{On spacetime symmetries\label{disc}}

It may be clear to readers that quite a different physical picture emerges from this model, as compared to the standard QFT.  This is largely due to the multiple independent $\mathfrak{so}(3)$ symmetries of equation~(\ref{5rot}) found in the quaternionic sector of the algebra.  We note the following:

\it 1.  At first glance, we find $\mathfrak{so}(3),$ not $\mathfrak{so}(3,1).$    \rm

It is perhaps a feature of this model that spatial symmetries enter the picture via equation~(\ref{5rot}) in exact analogy to internal symmetries.  

With this said, this does not mean that $\mathfrak{so}(3,1)$ is excluded from the model.  Lorentzian symmetry appears naturally, given that the proper orthochronous group, SO$^+$(3,1), comprises the non-trivial inner automorphisms of $\CH.$  (Furthermore, it should be pointed out that $\mathfrak{so}(3,1)$ provides the reduced structure algebra for $\mathcal{H}_2(\C).$)  Hence, in agreement with~\citep{Fotini},  \citep{thesis}, \citep{Fur2023}, \citep{sean} we propose that Lorentzian structure need not originate from a spacetime manifold.  Avoiding dependence on a spacetime manifold might prove fruitful in the context of quantum gravity~\citep{FR2022}, and could evade the inevitability of a block universe, which does not align with our experience of time.

\it 2.  The model exhibits a non-trivial relationship between generations (mass) and spin. \rm

As can be confirmed in Section~\ref{assemble}, this model represents an excitation from one  generation to another analogously to how it represents an excitation from one spin state to another.  With this said, these two properties are related to one another non-trivially.  That is, as mentioned above, $\CH$ is subject to inner automorphisms of the form $L_GR_{G^{-1}}$ for $G$ $\in$ SL(2,$\C$).  In practice, this means that  invertible operators in the generation action can effectively be transported into the spin action on the other side, and vice versa, via a judicious choice of $G$.  It would be of interest to investigate whether such a property could have implications for neutrino oscillations.

 3.  \it  There is an independent  $\mathfrak{so}(3)$ symmetry for each $s_j\in \mathcal{S}_n.$    \rm

In this model, we propose characterizing fermions as having two distinct endpoints ($\{s_i, s_j\}$ with $i\neq j$).  In this sense, we view a fermion as an extended object (representable as an edge) within the Jordan algebra.  It need not be embedded in any background spacetime.  

As mentioned earlier, we may think of spinors in this model as the ``square root" of $p_{\mu}\sigma^{\mu} = \sum \Psi \Psi^{\dagger}$ operators.  Of particular interest will be the usual symmetric cone associated with this Euclidean Jordan algebra, as it generalizes the forward light cone in 3+1 dimensions.  

This characterization of fermions, via asymmetric idempotents, allows for a possibility not available in most spinorial theories.  That is, the \it product of two distinct fermions may result in a third fermion. \rm  Such a property could be crucial in the description of three generations as a triality triple.  Unlike in~\citep{TTT}, all three generations could then admit a fermionic characterization.

On the other hand, gauge bosons have common endpoints ($\{s_i, s_i\}$), that is, before the theory is in fact gauged.  Upon the introduction of dynamics, however, one would anticipate a set of Peirce idempotents, $\mathcal{S}_n,$ that varies from point to point, as mentioned in the previous section.  

It is a future project to investigate if the usual spin statistics can be recovered in this setup.  Already one may see that any element in a  fermionic subspace $s_i \mathcal{H}_{16}(\C) s_j$ squares to zero.  On the other hand, any element within a boson subspace  $s_i \mathcal{H}_{16}(\C) s_i$ squares to another element within that same subspace.

In closing, we mentioned that the set of all fermion subspaces in this model may be represented by the complete graph,  $K_5.$  It would be interesting to see if such a structure could be connected to certain models of quantum computing, \citep{Hilary}, \citep{BoyleKulp}, \citep{LiBoyle},  \citep{Fotini}, \citep{seth}, \citep{MarCos}.

\begin{figure}[h]
\begin{center}
\includegraphics[width=7.5cm]{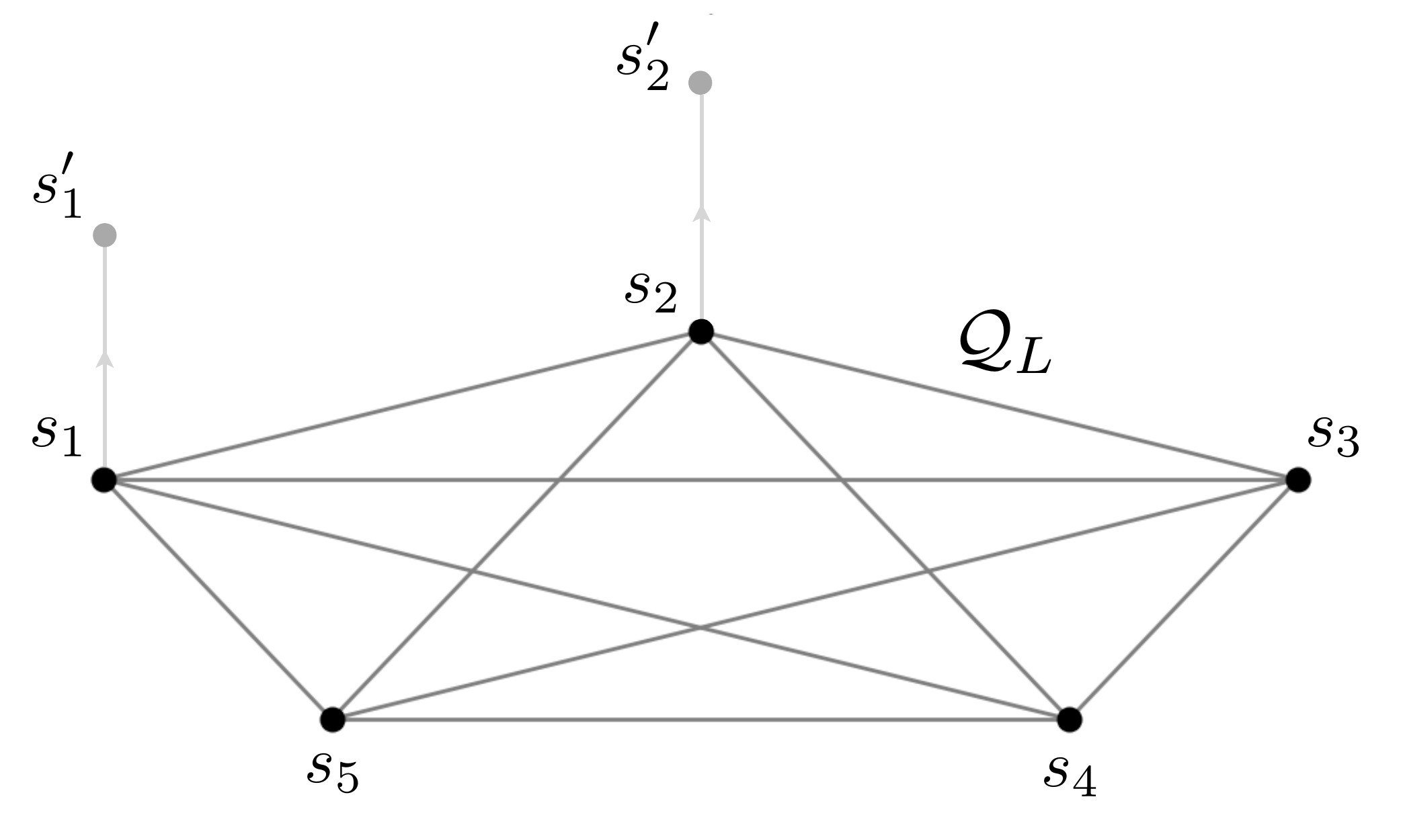}
\caption{\label{K5}  The complete graph, $K_5,$ representing fermions as extended objects. An independent $\mathfrak{so}(3)$ symmetry exists at each site $s_j$.  Dynamics induce subsequent Peirce idempotents $\{s_j'\}.$}
\end{center}\end{figure}

\section{Summary}

We extend the familiar $2\times2$ hermitian Weyl operator, $p_{\mu}\sigma^{\mu},$ to a new $16\times 16$ hermitian Weyl \it super \rm operator.  Written in the language of the normed division algebras, we define on $\mathcal{H}_{16}(\C)$ a $\mathbb{Z}_2^5$ grading, making it a superalgebra.  Octonionic internal symmetries respecting the grading are found to be $\mathfrak{u}(3)\oplus \mathfrak{u}(2)\oplus 3\hspace{.5mm}\mathfrak{u}(1)$.  In exact analogy, we find the remaining quaternionic spatial symmetries to be given by copies of $\mathfrak{su}(2) = \mathfrak{so}(3)$.

This grading elucidates a $64\hspace{.5mm}\mathbb{R}+3\cdot 64\hspace{.5mm}\mathbb{R}$ partitioning of the $256\hspace{.5mm}\mathbb{R}$-dimensional  $\mathcal{H}_{16}(\C).$   The first $64\hspace{.5mm}\mathbb{R}$ degrees of freedom accommodate gauge bosons, while the remaining  $3\cdot 64\hspace{.5mm}\mathbb{R}$ accommodate generations.

More explicitly, we find a $64\hspace{.5mm}\mathbb{R}$-dim momentum-space covariant derivative within the diagonal blocks, describing
$$\begin{array}{llllll} 
\color{gray}G_{\mu} &\left( \hspace{1mm} \underline{\mathbf{8}},\hspace{1mm} \underline{\mathbf{1}}, \hspace{1mm}0\hspace{1mm} \right)_4   &\hspace{4mm}  \color{gray}W_{L\mu} &\left( \hspace{1mm} \underline{\mathbf{1}},\hspace{1mm} \underline{\mathbf{3}}, \hspace{1mm}0\hspace{1mm} \right)_4   &  \hspace{4mm}\color{gray}B_{\mu} &\left( \hspace{1mm} \underline{\mathbf{1}},\hspace{1mm} \underline{\mathbf{1}}, \hspace{1mm}0\hspace{1mm} \right)_4 \vspace{2mm}\\

 \color{gray}p^{\ell}_{\mu}&\left( \hspace{1mm} \underline{\mathbf{1}},\hspace{1mm} \underline{\mathbf{1}}, \hspace{1mm}0\hspace{1mm} \right)_4 &\hspace{4mm}  \color{gray}p^{q}_{\mu}&\left( \hspace{1mm} \underline{\mathbf{1}},\hspace{1mm} \underline{\mathbf{1}}, \hspace{1mm}0\hspace{1mm} \right)_4  \  & \hspace{4mm} \color{gray}p^a_{\mu}&\left( \hspace{1mm} \underline{\mathbf{1}},\hspace{1mm} \underline{\mathbf{1}}, \hspace{1mm}0\hspace{1mm} \right)_4  \vspace{2mm}\\
 
 & &\hspace{4mm}  \color{gray}p^{\bar{a}}_{\mu}&\left( \hspace{1mm} \underline{\mathbf{1}},\hspace{1mm} \underline{\mathbf{1}}, \hspace{1mm}0\hspace{1mm} \right)_4,  \vspace{2mm}  & &
\end{array}$$
\noindent where the four $p_{\mu}\sigma^{\mu}$ relate to four $\mathfrak{u}(1)$ symmetries.  The subscripts here refer to the multiplicity.  In the outer off-diagonal blocks, we identify the complex representations for two generations, including sterile neutrinos, thereby accounting for another $2\cdot64\hspace{.5mm}\mathbb{R}$:

$$\begin{array}{rlrl} 
\color{gray}{(\hspace{1mm}  u, \hspace{1mm}d\hspace{1mm})_L} & \left( \hspace{1mm} \underline{\mathbf{3}},\hspace{1mm} \underline{\mathbf{2}}, \hspace{1mm}\frac{1}{6}\hspace{1mm} \right)_2 \hspace{6mm}   &  \color{gray} (\hspace{1mm}  \nu_e, \hspace{1mm}e\hspace{1mm})_L & \left( \hspace{1mm} \underline{\mathbf{1}},\hspace{1mm} \underline{\mathbf{2}}, \hspace{1mm}-\frac{1}{2}\hspace{1mm} \right)_2 \vspace{1mm}\\
\color{gray}(\hspace{1mm}  c, \hspace{1mm}s\hspace{1mm})_L & \left( \hspace{1mm} \underline{\mathbf{3}},\hspace{1mm} \underline{\mathbf{2}}, \hspace{1mm}\frac{1}{6}\hspace{1mm} \right)_2 \hspace{4mm}   &   \color{gray} (\hspace{1mm}  \nu_{\mu}, \hspace{1mm}\mu\hspace{1mm})_L & \left( \hspace{1mm} \underline{\mathbf{1}},\hspace{1mm} \underline{\mathbf{2}}, \hspace{1mm}-\frac{1}{2}\hspace{1mm} \right)_2  \vspace{1mm}\\

\color{gray}u_R & \left( \hspace{1mm} \underline{\mathbf{3}},\hspace{1mm} \underline{\mathbf{1}}, \hspace{1mm}\frac{2}{3}\hspace{1mm} \right)_2 \hspace{6mm}   &   \color{gray}\nu_{e_R} &\left( \hspace{1mm} \underline{\mathbf{1}},\hspace{1mm} \underline{\mathbf{1}}, \hspace{1mm}0\hspace{1mm} \right)_2  \vspace{1mm}\\

\color{gray}c_R & \left( \hspace{1mm} \underline{\mathbf{3}},\hspace{1mm} \underline{\mathbf{1}}, \hspace{1mm}\frac{2}{3}\hspace{1mm} \right)_2 \hspace{6mm}   &   \color{gray}\nu_{\mu_R} &\left( \hspace{1mm} \underline{\mathbf{1}},\hspace{1mm} \underline{\mathbf{1}}, \hspace{1mm}0\hspace{1mm} \right)_2  \vspace{1mm}\\

\color{gray}d_R & \left( \hspace{1mm} \underline{\mathbf{3}},\hspace{1mm} \underline{\mathbf{1}}, -\frac{1}{3}\hspace{1mm} \right)_2 \hspace{6mm}   &   \color{gray}e_{R} &\left( \hspace{1mm} \underline{\mathbf{1}},\hspace{1mm} \underline{\mathbf{1}}, -1\hspace{1mm} \right)_2  \vspace{1mm}\\

\color{gray}s_R & \left( \hspace{1mm} \underline{\mathbf{3}},\hspace{1mm} \underline{\mathbf{1}}, -\frac{1}{3}\hspace{1mm} \right)_2 \hspace{6mm}   &   \color{gray}\mu_R &\left( \hspace{1mm} \underline{\mathbf{1}},\hspace{1mm} \underline{\mathbf{1}}, -1\hspace{1mm} \right)_2.  \vspace{1mm}\\

\end{array}$$

Finally, in the inner off-diagonal blocks, we find the last $64\hspace{.5mm}\mathbb{R}$ degrees of freedom transforming as two copies of

$$\begin{array}{rlrl} 

\color{gray}b_R & \left( \hspace{1mm} \underline{\mathbf{3}},\hspace{1mm} \underline{\mathbf{1}}, -\frac{1}{3}\hspace{1mm} \right)_2 \hspace{6mm}   &   \color{gray} (\hspace{1mm}  \nu_{\tau}, \hspace{1mm}\tau\hspace{1mm})_L & \left( \hspace{1mm} \underline{\mathbf{1}},\hspace{1mm} \underline{\mathbf{2}}, \hspace{1mm}-\frac{1}{2}\hspace{1mm} \right)_2 \vspace{1mm}\\

\color{gray} h\cdot \overline{\nu_{\tau R}} & \left( \hspace{1mm} \underline{\mathbf{1}},\hspace{1mm} \underline{\mathbf{2}}, -\frac{1}{2}\hspace{1mm} \right)_2 \hspace{6mm}   &   \color{gray}  \tau_{R} & \left( \hspace{1mm} \underline{\mathbf{1}},\hspace{1mm} \underline{\mathbf{1}}, \hspace{1mm}-1\hspace{1mm} \right)_2, \vspace{1mm}

\end{array}$$
\noindent where we have taken the liberty to identify certain $\mathfrak{su}(2)_L$ doublets as Higgs-neutrino product states, $h\cdot \overline{\nu_{\tau R}}$.  Notably, this list excludes only those representations involving the top quark, most massive particle in the Standard Model. One possible explanation for this scenario could be that the top quark is in fact composite in nature, while the bottom quark is partially composite.  Anomaly cancellation will need to be addressed in future work.

\section{Outlook:  Bott Periodic Particle Physics and gravity \label{Bigpic}}

We close by describing current research directions.  First, we endeavour to understand how this current model fits into the larger framework of \it Bott Periodic Particle Physics, \rm first proposed in \citep{Gen}, \citep{Fur2021}, \citep{Fur2023}.  Extending our current left-multiplication model based on $L_{\A}$ to a full multiplication model based on $M_{\A}$ allows us to connect to  Bott periodicity.   It also enables  tetrad structure suitable for describing gravity.

\subsection{Bott Periodic Particle Physics}

When constructing a physical theory, perhaps the most difficult step is in deciding which fundamental mathematical object to choose.  On one hand, we might opt for a mathematical object that is somehow the \it most special. \rm Take for example, the unique exceptional Jordan algebra, $H_3(\mathbb{O}),$ or the Lie algebra $\mathfrak{e}_8,$ maximal within the exceptional simple Lie algebras.

Alternatively, we might opt for a mathematical object that is somehow the \it most common. \rm At least within the realm of non-degenerate Clifford algebras, $Cl(0,8)$ and $\CLtwo$ can be viewed as being the most common.  That is, Bott Periodicity tells us that the vast majority of  non-degenerate real Clifford algebras $Cl(p,q)$ can be decomposed as matrix algebras in the form:
\begin{equation}\begin{array}{ll}Cl(p,q) &= Cl(p_0, q_0)\otimes Cl(0,8) \otimes Cl(0,8)\otimes Cl(0,8) \dots \vspace{2mm}\\
&= Cl(p_0, q_0) \otimes Cl(0,8)^{\otimes m}
\end{array}\end{equation}
\noindent for some large m.    Here, $p_0= p \textup{ mod } 8$ and $q_0= q \textup{ mod } 8.$ That is, internally, these Clifford algebras are  \it overwhelmingly dominated \rm by factors of $Cl(0,8)$.

Similarly, the vast majority of  non-degenerate complex Clifford algebras $\C l(n)$ can be decomposed as matrix algebras in the form:
\begin{equation}\begin{array}{ll}\C l(n) &= \C l(n_0) \otimes \C l(2) \otimes \C l(2) \otimes \C l(2)  \dots \vspace{2mm}\\
&= \C l(n_0) \otimes \C l(2)^{\otimes m}
\end{array}\end{equation}
\noindent for some large m.  Here, $n_0 = n \textup{ mod } 2.$  Internally, these Clifford algebras are \it overwhelmingly dominated \rm by factors of $\C l(2)$.  Readers may appreciate that these $\C l(2)$ factors may be interpreted as  qubit algebras.

The Tenfold Way, \citep{10FWMoore}, \citep{10FWBaez}, is tied to these repeating structures, and, we speculate, may very well underlie elementary particle physics, \citep{Fur2023}.

Now, one motivation for studying the algebra $\RCHO$ came from the finding that $\CH\simeq \mathbb{R}\otimes\CH\simeq\CLtwo$ concisely describes each of the known Lorentz representations of the Standard Model, \citep{thesis}, \citep{Fur2023}.  Inclusion of the final normed division algebra, $\mathbb{O},$ provides a viable medium with which to generate internal symmetries.

However, $\A=\RCHO$ might also be useful for other purposes. \it Namely, it generates $Cl(0,8)\otimes\CLtwo.$ \rm Consider the full multiplication algebra of $\A$, denoted $M_{\A},$ earlier defined in Section~\ref{multalg}.  As matrix algebras, $M_{\A},$ begets a factorized form,  
\begin{equation}\begin{array}{ll} M_{\A} &\simeq \CLten \vspace{2mm}\\
&\simeq Cl(0,8)\otimes \CLtwo.
\end{array}\end{equation}
\noindent It is anticipated that this will be a recurring structure within the construction of a \it Bott Periodic Fock space, \rm as described in~\citep{Fur2023}.

\subsection{Incorporating gravity}

Beyond this, $M_{\A}$ has other useful decompositions, such as those leveraged in this article.  Explicitly,
\begin{equation}\begin{array}{rccc}
M_{\A} \simeq &\CLsix&\otimes_{\C}&\left(\CLtwo\otimes_{\C}\CLtwo\right).\vspace{2mm}\\
&\sim M_{\CO}&&\sim M_{\CH}
\end{array}\end{equation}
\noindent Under Lie-Jordan splitting, we have 
\begin{equation} M_{\A} \simeq \mathcal{H}_{32}(\C) \oplus \mathfrak{u}(32),
\end{equation}
\noindent whose hermitian part then decomposes as
\begin{equation} \mathcal{H}_{32}(\C)  \simeq \mathcal{H}_{8}(\C) \tilde{\otimes} \mathcal{H}_{2}(\C)\tilde{\otimes} \mathcal{H}_{2}(\C) ,
\end{equation}
\noindent where the symbol $\tilde{\otimes}$ is meant to remind readers that the product is defined non-trivially.

In this article, we have described something close to the Standard Model's particle spectrum within the first two factors of this product, 
\begin{equation} \mathcal{H}_{16}(\C)  \simeq \mathcal{H}_{8}(\C) \tilde{\otimes} \mathcal{H}_{2}(\C),
\end{equation}
\noindent with $\mathcal{H}_{16}(\C) $ given by the hermitian part of $L_{\A}\subset M_{\A}.$  $\mathcal{H}_{8}(\C)$ is octonionic, and describes internal degrees of freedom, while $\mathcal{H}_{2}(\C)$ is quaternionic, and describes spin/helicity.  These $\mathcal{H}_{16}(\C)\simeq \mathcal{H}_{L_{\A}}(\C)$ degrees of freedom are often associated with the fibre in the Standard Model, when viewed as a QFT.  

One may interpret $\mathcal{H}_{2}(\C)$ as a set of $4\hspace{.5mm}\mathbb{R}$ dimensional objects of the form $\sim p_{\mu}\sigma^{\mu},$ that is, with coefficients of a single spacetime index.  Extending this, one may identify the $\mathcal{H}_{2}(\C)\tilde{\otimes} \mathcal{H}_{2}(\C)$  as describing an object with two spacetime indices, although, with a non-trivial product $\tilde{\otimes}$ taken into account.  We propose identifying an element $e^{\mu}_{\nu}\in \mathcal{H}_{2}(\C)\tilde{\otimes} \mathcal{H}_{2}(\C)$ as a form of vierbein.  

From these beginnings, we aim to produce an emergent theory of gravity, free from an \it a priori \rm notion of spacetime,~\citep{thesis}.  The full Jordan algebra $\mathcal{H}_{M_{\A}}(\C)\simeq \mathcal{H}_{32}(\C)$ may be understood as an algebra of extended vierbeins for the Standard Model.

\section{Outlook:  $\OHCR$\label{plus}}

One long-standing challenge in this construction has been to find the origin of the quaternionic idempotents of our Peirce decomposition.  See Section~\ref{EW}.

Here we  propose a solution by reframing this model in terms of the 15$\hspace{.5mm}\R$ dimensional algebra 
$$\OHCR.$$
\noindent Specifically, one may consider the subalgebras of those multiplication algebras (endomorphisms) of $\mathbb{R},$ $\mathbb{C},$ $\mathbb{H},$ $\mathbb{O}$ commuting with a complex structure specified by the volume element of their corresponding Clifford algebras.  Specifically, 
\begin{equation} \begin{array}{lll} 
\textup{\bf Multiplication algebra}&&\textup{\bf Subalgebra}\vspace{3mm}\\
End_{\R}(\mathbb{O})\simeq Cl(0,6) \simeq Cl(4,2)  &\hspace{1mm}\mapsto  \hspace{1mm}&\C\oplus M_3(\C) \vspace{2mm}\\
End_{\R}( \mathbb{H}) \simeq Cl(3,1) &\hspace{1mm}\mapsto   \hspace{1mm}&M_2(\C) \vspace{2mm}\\
 End_{\R}( \mathbb{C}) \simeq Cl(2,0) &\hspace{1mm}\mapsto  \hspace{1mm}&\C\vspace{2mm}\\
  End_{\R}(  \mathbb{R})\simeq Cl(0,0) &\hspace{1mm} \mapsto  \hspace{1mm}&\R,
\end{array}\end{equation}
where we have taken care to account for both left- and right-multiplication algebras.  Doing so produces a set of diagonal blocks as $\C\oplus M_3(\C)\oplus M_2(\C)\oplus \C \oplus \R,$ in close connection with those of Figure~\ref{map}.   

It is important to note that the quaternionic case above differs significantly from the others, since it comprises the unique division algebra whose left- and right-multiplication algebras lead to distinct sets of endomorphisms.  As such, one may view its endomorphisms as $End_{\R}( \mathbb{H}) \simeq Cl(0,2)\otimes Cl(0,2).$  Reducing each factor to its even subalgebra gives
\begin{equation} \begin{array}{lll} 
End_{\R}( \mathbb{H}) \simeq Cl(0,2)\otimes Cl(0,2)&\hspace{1mm}\mapsto   \hspace{1mm}&\C\otimes \C \simeq \C\oplus \C. \vspace{2mm}\\
\end{array}\end{equation}
\noindent \it One may anticipate this fact to be the source of electroweak symmetry breaking. \rm

We  propose to study these structures in the context of  Connes' non-commutative geometry, \citep{Connes}, which is canonically based on the algebra $M_3(\C)\oplus \mathbb{H}\oplus \C$.

\section{Outlook:  $\mathbb{R}\subset \mathbb{C}\subset \mathbb{H}\subset \mathbb{O}\subset \mathbb{S}$  \label{sed}}

The research directions of \it Bott Periodic Particle Physics \rm and of $\OHCR$ \it modules \rm may in fact merge when this model is reframed in terms of the 16$\hspace{.5mm}\R$ dimensional sedenion algebra, $\mathbb{S}$.  It is known that $End_{\mathbb{R}}(\mathbb{S})\simeq Cl(0,8).$  Furthermore, $\mathbb{S}$ admits a vector space decomposition as 
\begin{equation} \mathbb{S} \mapsto e_8\mathbb{O} \oplus e_5\mathbb{H} \oplus  e_6\mathbb{C} \oplus e_7\mathbb{R} \oplus \mathbb{R}, 
\end{equation}
\noindent for $e_8 \in \mathbb{S}.$  

This may be seen to encode the nested inclusions $$\mathbb{R}\subset \mathbb{C}\subset \mathbb{H}\subset \mathbb{O}\subset \mathbb{S}$$ according to the Cayley-Dickson process.  

Repeating the procedure from the previous section leads to a set of diagonal blocks $\C\oplus M_3(\C)\oplus M_2(\C)\oplus \C \oplus \R \oplus \R.$  The particle content in this case has much in common with that of Figure~(\ref{map}).  Notably, one finds this time a counting suggestive of on-shell states.  (Transverse polarizations of gauge bosons may be enabled by the auxiliary imaginary units used in the Cayley-Dickson process.)  

The details in these final sections are to be carefully verified, and a complete bibliography of earlier related work is to be compiled.

\smallskip

\vspace{1.5cm}

\begin{acknowledgments}  

\it To my mother, and her sense of adventure. \rm

This article has benefitted from numerous discussions with Beth Romano and Mia Hughes.  The author is furthermore grateful for feedback from Brage Gording, Rob Klabbers, Jens K\"{o}plinger, Alex Nietner, Agostino Patella, Angnis Schmidt-May, Shadi Tahvildar-Zadeh, Graham White, Jorge Zanelli, and for the kind hospitality of Dariusz Chruscinski, Milosz Michalski, Gniewko Sarbicki, and the mathematical physics group at Nicolaus Copernicus University.  

This work was graciously supported by the VW Stiftung Freigeist Fellowship, Humboldt-Universit\"{a}t zu Berlin, and a visiting fellowship at the African Institute for Mathematical Sciences in Cape Town.


\end{acknowledgments}

\medskip

\end{document}